\documentclass[reprint,aps,amsmath,amssymb,superscriptaddress]{revtex4-1}

\usepackage{graphicx} 
\usepackage[utf8]{inputenc}
\usepackage{bm}
\usepackage{hyperref}
\usepackage{xcolor}
\usepackage[normalem]{ulem}
\usepackage{amsmath}
\usepackage{float}
\usepackage{algorithm}
\usepackage{algpseudocode}
\usepackage{comment}

\begin{document}

\title{Geometry of Urban Order}

\author{Marc Barthelemy}
\email{marc.barthelemy@ipht.fr}
\affiliation{Universit\'e Paris-Saclay, CNRS, CEA, Institut de Physique Th\'eorique,
91191 Gif-sur-Yvette, France}
\affiliation{Centre d'Analyse et de Math\'ematique Sociales (CNRS/EHESS), Paris, France}
\affiliation{Complexity Science Hub, Vienna, Austria}

\author{Geoff Boeing}
\email{boeing@usc.edu}
\affiliation{Department of Urban Planning and Spatial Analysis,
Sol Price School of Public Policy, University of Southern California,
301A Lewis Hall, Los Angeles, CA 90089-0626, USA}

\date{\today}

\begin{abstract}

Urban street networks are modeled as spatial graphs whose geometry reflects
planning, history, and physical constraints. Global orientation measures
describe the overall order of a city but do not reveal how street
directions form coherent neighborhoods. Here we introduce \emph{bearings
communities}---spatially connected domains whose streets share a common
orientation---and analyze them using concepts from the physics of
polycrystalline materials. Across $1025$ large cities worldwide, these
domains are almost perfectly ordered: their mean coherence is $0.964$
and varies by only $1.4\%$ from one city to the next, far above the
value expected for randomly oriented streets. Every city is therefore a
mosaic of well-ordered orientational grains, differing only in their
number, size, and mutual alignment. Because the grains are equally
regular everywhere, a city's global order is set mainly by how many of
them it averages over: a city appears ordered because it is built from a
few large grains, not because its fabric is locally more regular. A null
model in which grain orientations are randomized confirms this
grain-counting mechanism. Finally, the interface fraction $\gamma$
measures how strongly domains interlock, yielding a morphology diagram
that organizes cities continuously between grid-like, polycrystalline,
and fragmented limits.

\end{abstract}

\maketitle

\section{Introduction}

Urban street networks have been extensively studied over the past two decades through a complex spatial networks framework (see \cite{barthelemy2022spatial} and references therein). A wide range of structural, geometric, and dynamical properties have been investigated,
including connectivity patterns, centrality measures, scaling relations, cost--efficiency trade-offs, and typological classification \cite{Marshall:2006,Cardillo:2006,Buhl:2006,Xie:2007,Crucitti:2006,Lammer:2006,Porta:2006,Kalapala:2006,Barthelemy:2013,Strano:2012,Louf:2014,Levinson:2012,kirkley2018betweenness,chen2024global,gudmundsson2013entropy,Jiang:2007,Scellato:2006,Masucci:2009,Rosvall:2005,Strano:2013,lagesse2015spatial,Marshall:2018,Boeing:2019,Barthelemy2024Review,Louf:2014b,Thompson:2020,boeing_urban_2024,Chen:2024,batik2026best}.
Beyond static descriptions, the temporal evolution of street networks has also received increasing attention with the digitization of historical maps and infrastructure data,
revealing strong path dependence and long-term persistence in urban form
\cite{Perret:2015, Strano:2012, Masucci:2013, Barthelemy:2013, Burghardt:2022,
barrington2015century}.

Due to strong physical and spatial constraints, many quantities commonly studied in
complex networks \cite{latora2017complex, menczer2020first} have limited relevance for street networks.
Planarity and spatial embedding impose a narrow degree distribution and a high clustering coefficient, largely fixed by geometry
\cite{Lammer:2006,barthelemy2022spatial}.
Similarly, several classical indicators from transportation geography
\cite{Kansky:1963} depend mainly on the average
degree and therefore provide little independent information for large urban networks. As a result, effective characterization of street networks requires spatially-informed measures that directly reflect geometric and organizational  \cite{Barthelemy2024Review}.
These include planarity, the distribution of road segment lengths, the spatial distribution of betweenness centrality, block shape factors, and the distribution of street orientations. In particular, street orientation has emerged as a simple interpretable signature of urban morphology \cite{gudmundsson2013entropy,Boeing:2019}, capturing the contrast between planned grid-like layouts and incrementally evolved fabrics.
A major contribution in this direction analyzed street network orientation and entropy across 100 cities worldwide, demonstrating that orientation-based measures reveal systematic differences between urban systems and correlate with circuity,
connectedness, and efficiency \cite{Boeing:2019}. However, orientation entropy and related global indicators average over space and do
not capture how street orientations organize into spatially connected neighborhoods,
nor how adjacent neighborhoods connect through their interfaces. Consequently, cities with markedly different internal structures---such as variations of a mosaic of rotated grids---would produce the same high orientation entropy value, which gives no information on their internal differences.

This limitation calls for a mesoscopic description of urban street networks,
intermediate between global indicators and local geometry.
Many cities are best described as patchworks of districts reflecting distinct planning epochs, topographical constraints, or growth processes, suggesting an analogy with orientational systems spanning coherent, polycrystalline, and fragmented regimes. Here, we develop such a mesoscopic framework---explicitly identifying orientational domains and their interfaces, in analogy with grains and grain boundaries in polycrystalline materials \cite{priester2012grain,sutton1995interfaces}. The logic of the paper is then the following: we first decompose each
city into orientational domains and characterize them individually,  we then introduce a null model of a random mosaic, and finally, 
we place all cities in a
two-dimensional morphology diagram whose limiting regimes are the
urban analogues of single crystals, polycrystals, and fine-grained
fragmented fabrics.

\section{Bearings communities}


\begin{figure}
	\centering
     \includegraphics[width=0.49\textwidth]{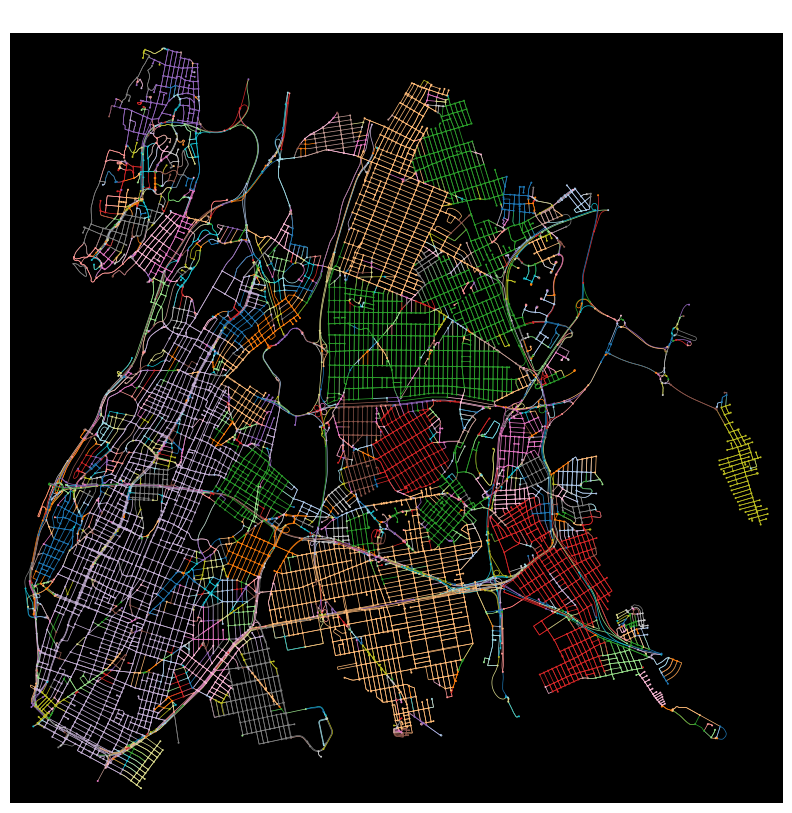}
     \caption{Illustration of the decomposition of a street network into
bearings communities---spatially connected domains sharing a common
street orientation---for The Bronx (United States). Each color denotes
a distinct community identified by the algorithm described in the
text: several medium-sized, internally ordered domains with different
dominant orientations tile the network, meeting along sharp
interfaces. Drivable street network from
OpenStreetMap~\cite{OSM}, angular tolerance
$\delta = 10^\circ$.}
	\label{fig:example}
\end{figure}

We represent a street network as a planar spatial graph, where edges
correspond to street segments and nodes to intersections. For each city
we analyze the drivable network within a fixed disc of radius $10$~km
centered on the city, extracted from OpenStreetMap \cite{OSM}; using an
identical spatial window for all cities ensures that the measures
defined below are directly comparable across the dataset (see section I of the SM for
details on data and preprocessing). Each edge~$e$
is characterised by its length~$\ell_e$ and its bearing
$\theta_e\in[0,\pi)$, defined modulo~$\pi$ to enforce head-tail
symmetry. \emph{Bearings communities} are defined as spatially connected
regions sharing a common local orientational structure. The construction
is performed at the node level: for each node~$n$, we consider the
bearings $\{\theta_{nj}\}_{j\sim n}$ of all incident edges and define a
dominant orientation via the circular mean on the $[0,\pi/2)$ interval,
\begin{equation}
\phi_n \;=\; \frac{1}{4}\,\arg\!\left\langle\,
e^{\,4\mathrm{i}\,\theta_{nj}} \,\right\rangle_{j\sim n}
\;\bmod\;\tfrac{\pi}{2}\,,
\label{eq:def_phi}
\end{equation}
so that $\phi_n\in[0,\pi/2)$ (the angle brackets
denote the arithmetic mean over all incident edges). The factor of~$4$ in the exponent maps
the fourfold-symmetric orientations onto the full circle, ensuring that
angles near opposite ends of the $[0,\pi/2)$ interval---such as
$1^{\circ}$ and $89^{\circ}$---are correctly recognised as nearly
equivalent. This circular averaging avoids the spurious values that a
linear statistic would produce at the wrapping boundary, while naturally
reflecting the fourfold symmetry of grid-like patterns. Physically,
$\phi_n$ represents the locally preferred street axis at an intersection. Note that $\phi_n$ is undefined when $R_n = |\langle e^{4i\theta}\rangle_n|
\to 0$ (for instance at perfectly symmetric three-fold junctions); nodes with $R_n < R_{\min}$ (we took $R_{\min}=0.1$, see SM Sec. I) are therefore excluded from the orientational analysis.
Two adjacent nodes~$n$ and~$m$ are considered orientationally coherent if
\begin{equation}
d_{\pi/2}(\phi_n,\phi_m) \;<\; \delta\,,
\end{equation}
where $d_{\pi/2}(\alpha,\beta)
= \min\!\bigl(|\alpha-\beta|,\,\tfrac{\pi}{2}-|\alpha-\beta|\bigr)$
is the angular distance modulo~$\pi/2$, and $\delta$ is a fixed angular
tolerance (see the section I of the Supplementary Material for details). A bearings
community is then defined as a maximal connected set of nodes satisfying
this compatibility condition, identified by a seed-based region-growing
algorithm. Starting from an unassigned seed node, the algorithm visits
graph-neighbours and adds those whose dominant orientation lies within
$\delta$ of the \emph{seed} orientation (this definition does not require nodes of degree four: each node is assigned a local dominant orientation from its adjacent street segments, whatever its degree); the process iterates until no
further compatible neighbours can be included. Comparing every
candidate to the seed rather than to its immediate predecessor prevents
slow orientational drift across space, playing a role analogous to a
misorientation tolerance in polycrystalline materials and ensuring that
communities remain compact and internally coherent. The resulting partition decomposes the network into contiguous
orientational domains---the bearings communities.

Figure~\ref{fig:example} illustrates this decomposition for a single
city, The Bronx (United States); the data here and in the following
come from OpenStreetMap \cite{OSM}. The network decomposes into
several medium-sized communities, each internally well ordered but
with a dominant orientation differing from that of its neighbors, so
that adjacent domains meet along sharp orientational interfaces. This
example displays the generic structure exploited throughout
this work: bearings communities behave as orientational ``grains'',
separated by interfaces where the dominant street direction changes
abruptly, providing a mesoscopic description that bridges local
geometry and global urban organization.

\section{Measures}

\subsection{Cluster misorientation}



To characterize how orientational domains meet, we define the misorientation angle between two neighboring bearings communities $c$ and $c'$ in analogy with the theory of
grain boundaries in polycrystalline materials
\cite{read1950dislocation,sutton1995interfaces},
\begin{equation}
\Delta\phi = \min\!\left(\,\left|\phi_c-\phi_{c'}\right|,\;
\frac{\pi}{2}-\left|\phi_c-\phi_{c'}\right|\right),
\label{eq:misor}
\end{equation}
where $\phi_c,\phi_{c'}\in[0,\pi/2)$, so that $0\le\Delta\phi\le\pi/4$,
the maximum being attained when the two domains are rotated by
$45^\circ$ with respect to one another. Here $\phi_c$ denotes the dominant orientation of community $c$, evaluated as the
circular mean of the node orientations $\phi_n$ over the community,
which weights nodes equally rather than weighting edges by their length (as in Eq.~(\ref{eq:zc}) below).
\begin{figure}
	\centering
     \includegraphics[width=0.49\textwidth]{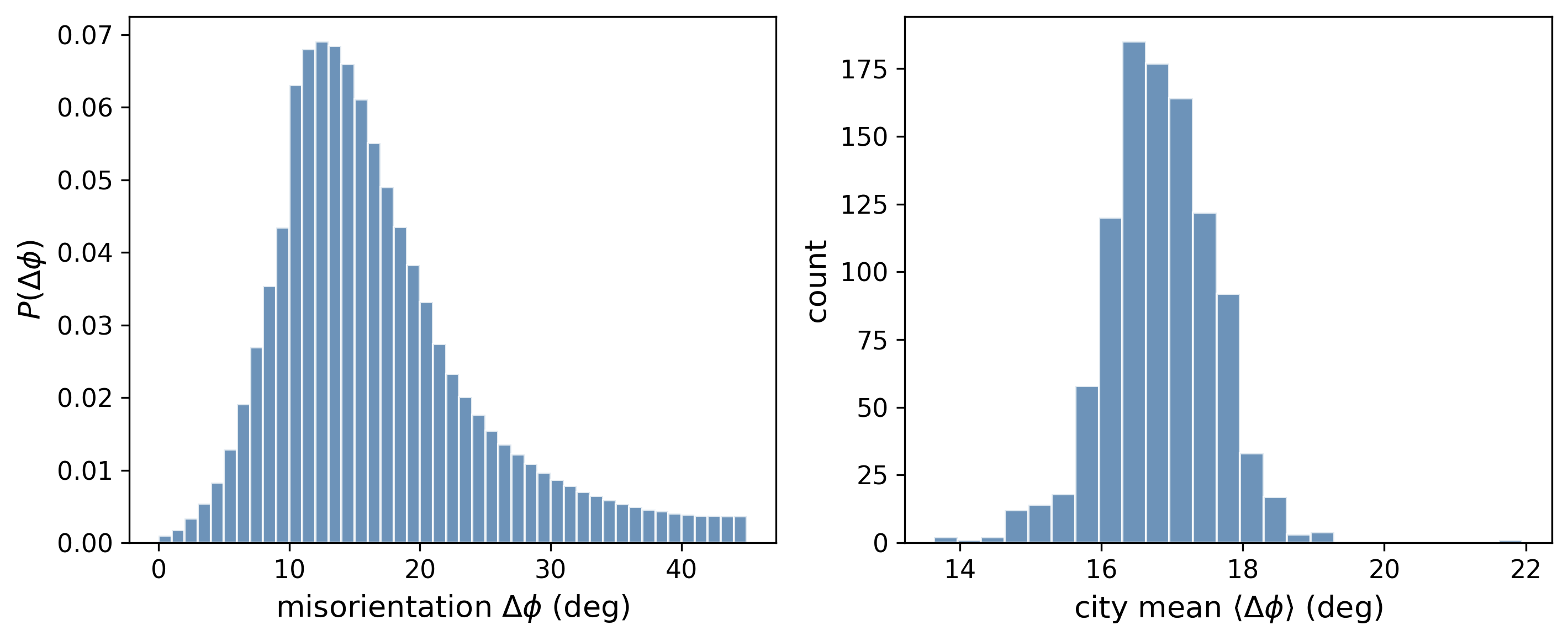}
	\caption{
(Left) Distribution of the misorientation angle $\Delta\phi$ over all interfaces between neighboring bearings communities, aggregated across the $1025$ cities (obtained for $\delta=10^\circ$). (Right) Distribution of the city-level average misorientation $\langle \Delta\phi \rangle$, highlighting a characteristic mesoscopic mismatch between orientational domains.
}
	\label{fig:deltatheta}
\end{figure}

The distribution $P(\Delta\phi)$, computed over all inter-community interfaces and weighted by interface length, is broad but biased toward relatively small misorientations above the clustering threshold $\delta$ (Fig.~\ref{fig:deltatheta}). Its mean and standard deviation are $16.8^\circ$ and $8^\circ$, respectively. Thus, misorientation angles vary substantially from one interface to another, with no sharply selected local angle. By contrast, the city-level average misorientation $\langle\Delta\phi\rangle$ is narrowly distributed across the $1025$ cities, each city is therefore characterized by a well-defined mesoscopic angular separation between its orientational domains (see section II of the SM for more details).

\subsection{Interface analysis}

To quantify the spatial fragmentation of orientational domains, we introduce an interface measure defined at the level of bearings communities. For each community $c$, we define the total internal street length
\begin{equation}
L_c = \sum_{(i,j)\in E} \ell_{ij}\,\mathbf{1}(c_i=c_j=c),
\label{eq:Lc}
\end{equation}
and the total interfacial street length
\begin{equation}
\Lambda_c = \sum_{(i,j)\in E} \ell_{ij}\,\mathbf{1}(c_i=c,\; c_j\neq c),
\end{equation}
where $E$ is the set of edges of the network, $\ell_{ij}$ is the physical length of edge $(i,j)$, and $c_i$ denotes the bearings
community of node $i$. We then define the cluster-level interface density as $\gamma_c = \Lambda_c/L_c$, which compares the boundary streets of community $c$ to its internal street length.
Small values of $\gamma_c$ correspond to compact, well-insulated orientational domains, whereas large values indicate highly perforated or strongly interwoven clusters. To obtain a city-level scalar, we compute the length-weighted average
\begin{equation}
\gamma = \frac{\sum_c L_c\,\gamma_c}{\sum_c L_c}
= \frac{\sum_c \Lambda_c}{\sum_c L_c},
\end{equation}
which represents the typical boundary-to-bulk ratio experienced within the network---in the language of materials, a surface-to-volume ratio of the grain structure. Because each interface separates two communities, it contributes to
$\Lambda_c$ for both, so that $\sum_c\Lambda_c$ counts each interface
segment twice: $\gamma$ is the boundary-to-bulk ratio seen from either
side of an interface. It is related to the global interface fraction
$\Gamma=I/L$, where $I$ is the total interface length counted once, by
$\gamma=2\Gamma/(1-\Gamma)$; see Sec.~III of the Supplemental Material.

This interface ratio $\gamma$ varies substantially across cities
(see Fig.~S3 in Sec.~III of the Supplemental Material). Across the
$1025$ cities, we find
$\langle \gamma \rangle = 0.63 \pm 0.20$, with a median of $0.611$. Values span from
$0.18$ to $3.04$, reflecting the wide diversity of
bearings-community shapes and degrees of fragmentation.

\subsection{Orientational order: the $Q_4$ parameter}

To quantify global orientational order in a planar street network, we introduce a family of generalized $m$-fold orientational order parameters
\begin{equation}
Q_m = \left| \frac{1}{L} \sum_e \ell_e\, e^{i m \theta_e} \right|
\label{eq:Qm}
\end{equation}
where $L = \sum_e \ell_e$, $\theta_e \in [0,\pi)$ denotes the bearing of edge $e$, and $\ell_e$ is its length. The integer $m$ selects the symmetry of interest, and $Q_m$ is invariant under rotations by an angle $2\pi/m$.
Such bond-orientational order parameters have been extensively studied in condensed
matter systems \cite{Strandburg2012BOO}. By construction, $Q_m$ depends only on the distribution of edge orientations and
is entirely independent of any clustering or partitioning of the network into bearings communities.

The case $m=2$ corresponds to the standard nematic order parameter (taking into account the head-tail symmetry), widely used in liquid crystals and active matter, and measures uniaxial alignment \cite{ChaikinLubensky}.
In urban networks, $Q_2$ vanishes for perfectly orthogonal grids: contributions from streets
oriented along perpendicular directions cancel exactly, reflecting the fact that nematic order is blind to multi-axis symmetry. To capture the dominant symmetry of orthogonal street layouts, we therefore focus on
the four-fold orientational order parameter $Q_4$
\begin{equation}
Q_4 = \left| \frac{1}{L} \sum_e \ell_e\, e^{4i\theta_e} \right|
\label{eq:Q4}
\end{equation}
For $m=4$, orientations separated by $\pi/2$ contribute constructively, so that perpendicular streets reinforce rather than cancel.
The quantity $Q_4$ thus measures the global coherence of orthogonal order in the network, independently of the absolute orientation of the grid: $Q_4=1$ for a perfect grid, whatever its orientation, and $Q_4=0$ for uniformly distributed bearings.

The identification of orientations separated by $\pi/2$ that makes
$Q_4$ useful also makes it degenerate in one respect: a bundle of
mutually parallel streets and a perfect two-axis grid both give
$Q_4=1$, and no fourfold quantity can tell them apart. Resolving the
ambiguity requires the nematic parameter $Q_2$ of the same
family~(\ref{eq:Qm}). Since we shall be concerned with order at the level of individual
bearings communities, we introduce the domain-level counterpart of
Eq.~(\ref{eq:Qm}). For a community $c$, let $\mathrm{int}(c)$ denote its
interior edges---those with both endpoints in $c$---and
$L_c=\sum_{e\in\mathrm{int}(c)}\ell_e$ its internal street length, as in
Eq.~(\ref{eq:Lc}). We then define
\begin{equation}
q_c^{(m)} \;=\; \left|\frac{1}{L_c}
\sum_{e\,\in\,\mathrm{int}(c)} \ell_e\, e^{im\theta_e}\right| ,
\label{eq:qmc}
\end{equation}
so that $q_c^{(2)}$ is the nematic order of the domain and $q_c^{(4)}$
its fourfold order; the latter plays a central role below and is
abbreviated $q_c$.

If a fraction
$f$ of the internal length lies along one axis and $1-f$ along the
perpendicular one, the two contributions to~(\ref{eq:qmc}) (for $m=2$) have
opposite signs and
\begin{equation}
q_c^{(2)} = \left|2f-1\right| ,
\label{eq:q2f}
\end{equation}
so that $q_c^{(2)}=0$ for a balanced grid ($f=1/2$) and $q_c^{(2)}\to1$ for
uniaxial alignment ($f=1$). Intermediate values measure the anisotropy of the blocks rather than the
absence of a second axis. For a rectangular grid, the streets running
along one direction are stacked across the perpendicular one, so their
total length is inversely proportional to their mutual spacing: blocks
of dimensions $d_x\times d_y$ carry street lengths $L_x\propto 1/d_y$
and $L_y\propto 1/d_x$, whence $L_x/L_y=d_x/d_y\equiv a$, giving $f=L_x/(L_x+L_y)=a/(1+a)$ leading to
\begin{equation}
q_c^{(2)} = \frac{a-1}{a+1} .
\end{equation}
The long axis of a block is thus also its dominant street direction: for
Manhattan's typically $80\,\mathrm{m}\times274\,\mathrm{m}$ blocks, $a\simeq3.4$
and $q_c^{(2)}\simeq0.55$.

Note that $q_c^{(2)}$ is meaningful only
domain by domain---a city-wide $Q_2$ would vanish trivially, since
neighbouring domains have unrelated axes whose nematic contributions
cancel.

Applied to the $1025$ cities, the length-weighted median is
$q_c^{(2)}=0.22$, corresponding to a typical block aspect ratio
$a\simeq1.6$. The distribution is strongly size-dependent. Domains with
$q_c^{(2)}>0.9$ carry $14.4\%$ of the total street length, but have a
median of only five interior edges and are trees in $99.5\%$ of cases: a tree contains no cycle and hence no block, so $q_c^{(2)}$ there measures the alignment of a few segments of a single street rather than the balance of two lattice axes. Restricting to domains large enough to contain blocks (interior edge
count $N_c\ge30$, defined in Sec.~\ref{sec:composites} below; these
carry $63.6\%$ of the street length), the length-weighted median falls to $q_c^{(2)}=0.13$---an aspect ratio
$a\simeq1.3$---and only $2.0\%$ of that length is uniaxial. The
fourfold domains that carry the urban fabric are therefore genuine
two-axis grids with mildly anisotropic blocks, not bundles of parallel
streets. Size-resolved statistics are given in Sec. IV of the Supplemental Material.

Because $Q_4$ is a global average, cities composed of many locally grid-like neighbourhoods with different orientations can exhibit a small $Q_4$ due to angular cancellation between domains---a direct analogue of
the vanishing net magnetisation of a polycrystal built from fully magnetized grains. The dominant node orientation $\phi_n$ (Eq.~\ref{eq:def_phi}) is built from the same
complex exponential $e^{4\mathrm{i}\theta}$ but extracts complementary
information: $\phi_n$ retains the \emph{phase} (the locally preferred
axis), whereas $Q_4$ retains the \emph{magnitude} (the degree of global
coherence).

\section{Cities as orientational composites}
\label{sec:composites}

\subsection{Grains and interfaces}

The measures introduced so far are of two kinds: $Q_m$ is a global
average over all edges, independent of the partition, while $\gamma$ and
$\Delta\phi$ describe the partition without referring to the order
inside each domain. To resolve the structure of the mosaic itself we
return to the domain-level parameter of Eq.~(\ref{eq:qmc}) at $m=4$ and
retain its phase as well as its modulus,
\begin{equation}
z_c \;=\; \frac{1}{L_c}\sum_{e\,\in\,\mathrm{int}(c)} \ell_e\, e^{4i\theta_e}
\;=\; q_c\, e^{4i\phi_c},
\label{eq:zc}
\end{equation}
where $q_c = |z_c| = q_c^{(4)} \in [0,1]$ measures how internally
ordered the domain is ($q_c = 1$ for a perfect grid, $q_c = 0$ for
random bearings) and $\phi_c = \tfrac14 \arg z_c$ is its dominant
orientation. Each domain is thus an arrow in the complex
plane---a \emph{phasor}---of length $q_c$ and direction $4\phi_c$.

\begin{figure}
	\centering
     \includegraphics[width=0.49\textwidth]{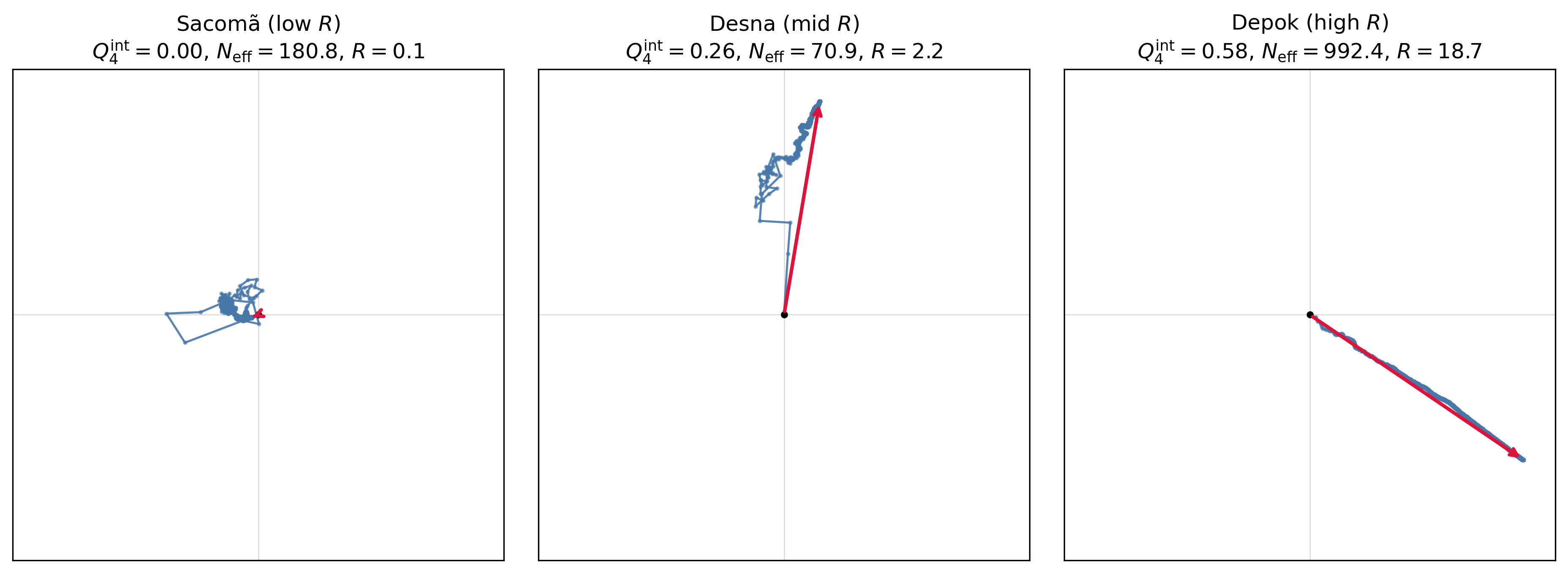}
	\caption{The phasor walk of Eq.~(\ref{eq:decomposition}) for three
	cities. Each arrow is one bearings community, of length $w_c q_c$
	(its street-length weight times its internal coherence) and
	direction $4\phi_c$ (four times its dominant orientation); arrows
	are drawn head-to-tail in order of decreasing weight, and the black
	resultant is the city's interior order parameter $Q_4^{\mathrm{int}}
	= |Z|$. Left to right: a fabric of misaligned domains, whose walk goes back to the origin and  leaves a small resultant (Sacom\~a, Brazil); an intermediate case (Desn\'a, Czech Republic); and a near-grid city (Depok, Indonesia), whose aligned steps produce a long, nearly straight walk resulting in a large value of $|Z|$.
    }
	\label{fig:phasorwalk}
\end{figure}

We now introduce the sum over all clusters of the grain-level coherence
\begin{equation}
Z \;\equiv\; \sum_c w_c\, z_c \;=\; \sum_c w_c\, q_c\, e^{4i\phi_c},
\label{eq:decomposition}
\end{equation}
where the weight $w_c = L_c/\sum_{c'}L_{c'}$ is the length fraction carried by
domain $c$. This is the central object of what follows: $Z$ is the
end-to-end vector of a ``phasor walk'' whose $c$-th step has length
$w_c q_c$ and direction $4\phi_c$ (Fig.~\ref{fig:phasorwalk}). The internal orientational order is therefore measured by
\begin{align}
Q_4^{\mathrm{int}} \equiv |Z|.
\end{align}
When the domains share a common orientation, the corresponding steps align and the walk has a large resultant, yielding a large $Q_4^{\mathrm{int}}$. When their orientations are scattered, the walk partly comes back to its origin and the resultant is small. 

This interior order parameter differs from the full-network $Q_4$ of
Eq.~(\ref{eq:Q4}) in two ways: it omits the interface edges, which join
two different communities and belong to no domain, and it is restricted
to communities of at least five nodes (Sec.~I of the Supplemental
Material). Empirically the two remain nearly proportional across the
dataset, $Q_4 = 0.85\,Q_4^{\mathrm{int}}$ (fit through the origin,
uncentred $R^2=0.995$, see Fig.~S2, section 3 of the SM). The two measures are thus interchangeable up to a constant
factor, and we use $Q_4^{\mathrm{int}}$ where the domain decomposition is
required---in the null model below---and the directly observable $Q_4$
elsewhere.

The triangle inequality bounds the walk by its total length,
\begin{equation}
Q_4^{\mathrm{int}} \;\leq\; \sum_c w_c\, q_c \;\equiv\; \bar q_{\rm city},
\label{eq:qbar_bound}
\end{equation}
with equality only when all domains are perfectly aligned; the
mean domain coherence $\bar q_{\rm city}$ is the order the city
\emph{would} have if its grains were all oriented the same way.

Two important facts about this decomposition organize the rest of the
paper. The first concerns the \emph{lengths} of the steps, and the
second their \emph{directions}.

Both the grain-level quantity $z_c$ and the city-level quantity $Z$ are
weighted sums of phasors. It is therefore useful to introduce a general
measure of the number of terms that effectively contribute to such a
sum. Consider
\begin{equation}
W=\sum_i a_i e^{i\alpha_i},
\qquad
a_i\geq 0,
\qquad
\sum_i a_i=1.
\label{eq:generic_phasor}
\end{equation}
If the phases $\alpha_i$ are independent and uniformly distributed,
the cross terms vanish on average, and
\begin{equation}
\left\langle |W|^2\right\rangle =\sum_i a_i^2.
\label{eq:weighted_phasor_variance}
\end{equation}
This motivates the participation number
\begin{equation}
N_{\rm part}[\{a_i\}]
\equiv
\frac{1}{\sum_i a_i^2},
\label{eq:general_participation}
\end{equation}
which measures the effective number of contributions to the sum. 
For $N$ equally weighted terms, $a_i=1/N$ and hence
$N_{\rm part}=N$. When a few
terms carry most of the weight, $N_{\rm part}$ is smaller than the
actual number of terms. This quantity is related to the inverse participation ratio familiar from
localization theory \cite{kramer1993localization}, or the $Y_2$ used to
count the largest pieces in fragmentation processes
\cite{Derrida:1987}.\\

\paragraph{Grains are internally ordered, in every city.}

\begin{figure}
\centering
\includegraphics[width=0.49\textwidth]{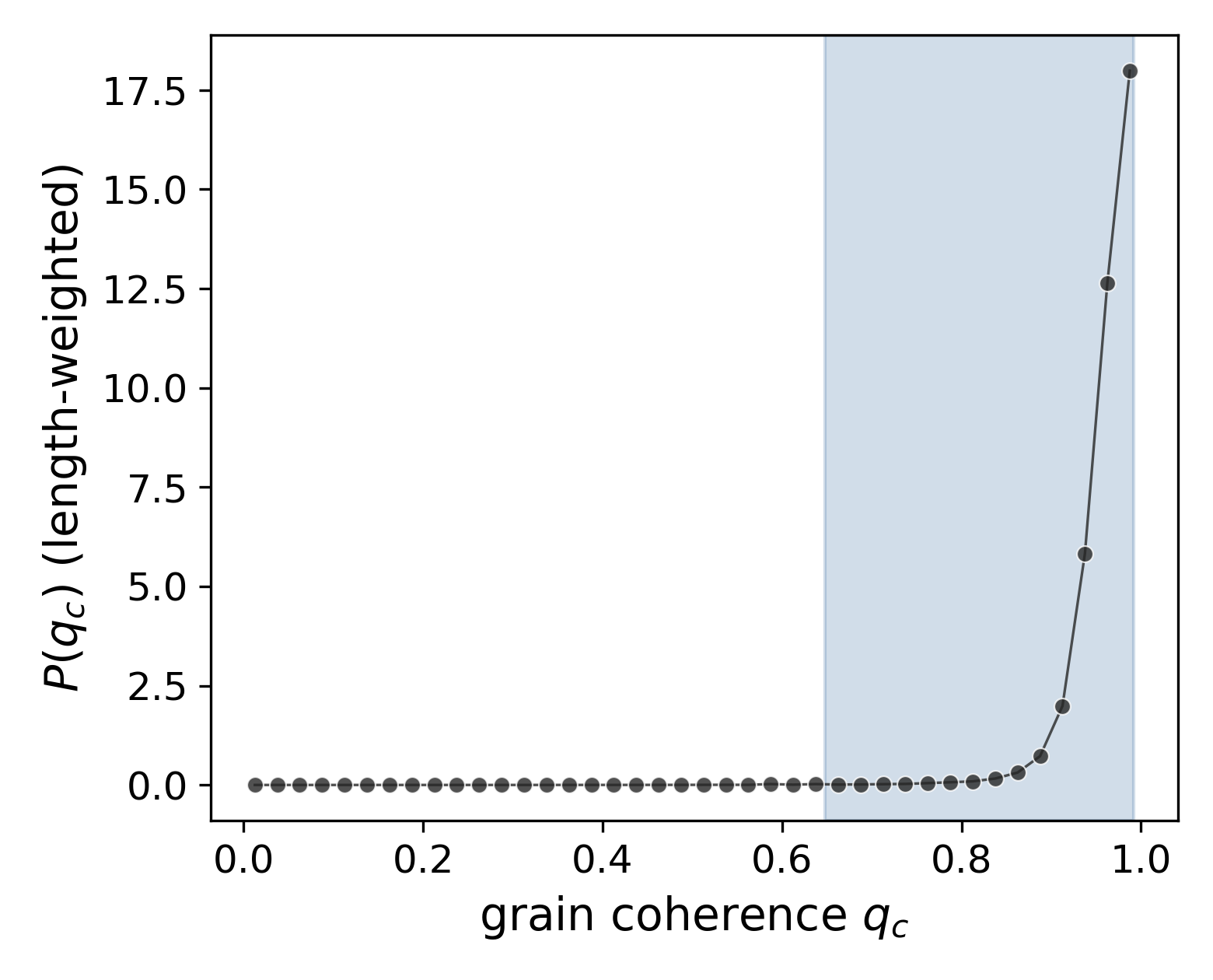}
\hfill
\includegraphics[width=0.49\textwidth]{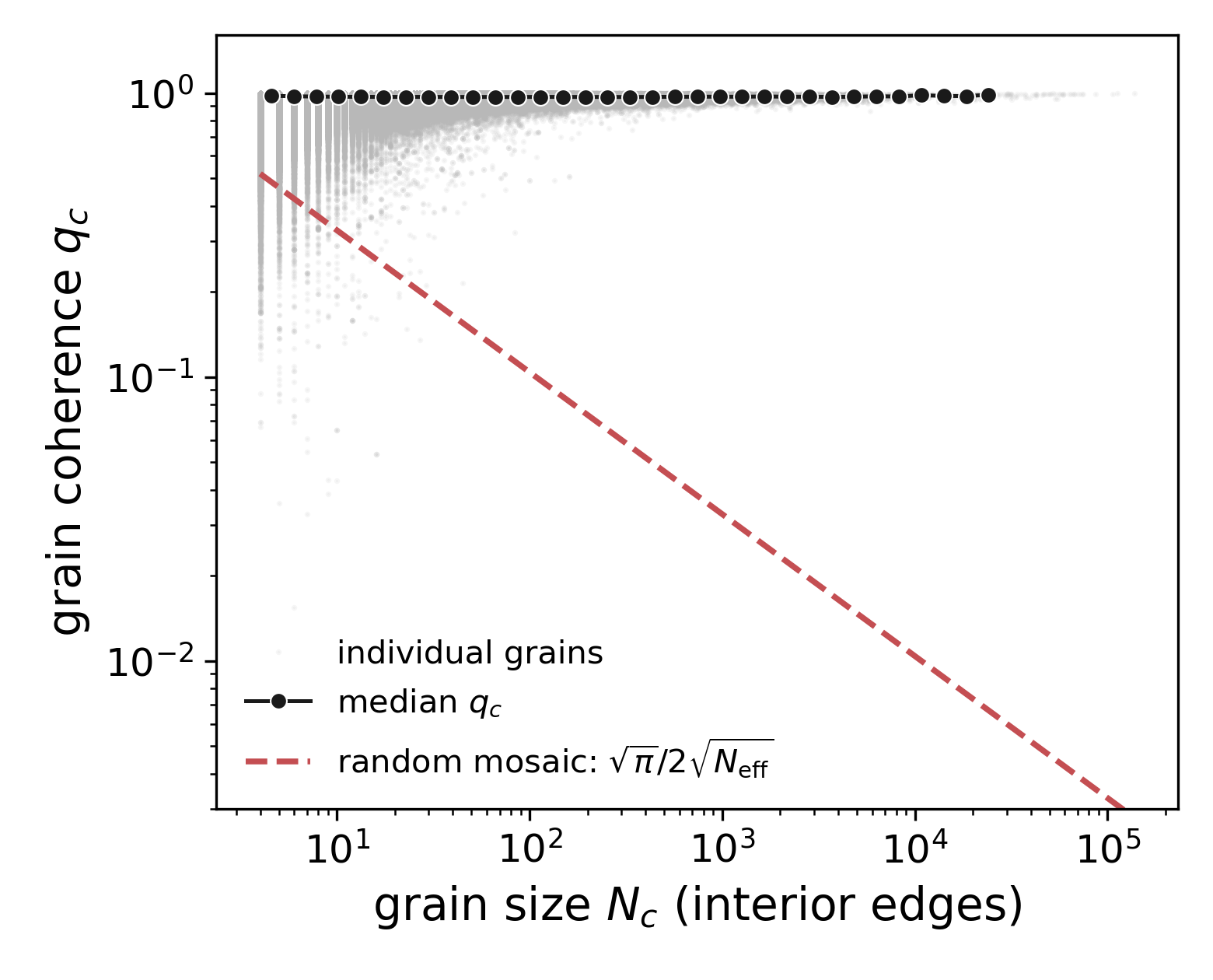}
\caption{
Grains are internally ordered in every city, at every scale.
\textit{Top:} Length-weighted distribution of the domain coherence
$q_c$ (Eq.~\ref{eq:zc}), pooled over all bearings communities with
at least $5$ nodes across the $1025$ cities. Almost no street length
lies in a poorly ordered domain ($q_c\lesssim0.8$); the distribution
is sharply peaked near unity, independently of the structural
regime of the parent city. The shaded band marks the $5$--$95\%$
range of the per-city mean coherence
$\bar q_{\rm city}=\sum_cw_cq_c$ across the dataset
(mean $0.964$, standard deviation $0.014$).
\textit{Bottom:} Domain coherence versus domain size $N_c$, defined
as the number of interior edges. Grey points represent individual
domains ($2.84\times10^6$ in total), and black symbols show the
median $q_c$ in logarithmic bins of $N_c$. The dashed line is the
random-bearing expectation of Eq.~(\ref{eq:qrand}), evaluated using
the median relation
$n_{\rm eff}^{(c)}/N_c=0.73$. The measured coherence is essentially
independent of size, varying only between $0.969$ and $0.977$ over
five decades, while the random baseline decreases as
$\left(n_{\rm eff}^{(c)}\right)^{-1/2}$. The excess coherence over
chance therefore increases with domain size.
}
\label{fig:composite}
\end{figure}

The step lengths are set by the domain coherences $q_c$, and these are
close to their maximum almost everywhere. Pooling all domains across the
$1025$ cities (Fig.~\ref{fig:composite}, top), essentially no street
length belongs to a poorly ordered domain: $96.1\%$ of the street length
lies in domains with $q_c>0.9$, and even the first percentile of the
length-weighted distribution reaches $q_c=0.833$. Equivalently, the
per-city mean coherence $\bar q_{\rm city}$
varies little from one city to the next: across the dataset $\left\langle \bar q_{\rm city}\right\rangle = 0.964\pm0.014$. All averages over domains are weighted by street length, since the
typical domain is small and carries a negligible share of the urban
fabric.

Three quantities should be distinguished: the coherence $q_c$ of a
single domain, the length-weighted mean
$\bar q_{\rm city}$ over the domains of one city, and its
average $\langle\bar q_{\rm city}\rangle$ over the $1025$ cities.
Angular brackets denote an average across the dataset unless otherwise
specified. It is $\bar q_{\rm city}$ that enters the city-level null
model below, while the small dispersion of
$\bar q_{\rm city}$---only $1.4\%$ of its mean---shows
that the internal order of the grains is nearly universal across
cities.

This high coherence is not imposed by the clustering procedure. The
tolerance $\delta$ is applied only to the dominant orientation
$\phi_n$ of each node (Eq.~\ref{eq:def_phi}), obtained by averaging the
bearings of the edges incident to that node. By contrast, the coherence
$q_c$ of a domain is computed directly from the bearings $\theta_e$ of
all its individual edges. Grouping nodes with similar average
orientations therefore does not require all their edges to be similarly
oriented.

For example, a node may have a dominant orientation close to the
domain axis while still carrying edges pointing in substantially
different directions. In the fourfold phasor sum defining $q_c$, an
edge oriented at $45^\circ$ relative to the domain axis contributes in
the direction opposite to that of an aligned edge. The clustering rule
therefore does not prevent a domain from having a small $q_c$. Its high
internal coherence is an empirical result, rather than a direct
consequence of the tolerance $\delta$.

The appropriate reference is therefore a domain whose individual edge
bearings are independent and random. Within domain $c$, define the
normalized length carried by edge $e$ as
\begin{equation}
a_e^{(c)}=\frac{\ell_e}{L_c},
\qquad
L_c=\sum_{e\in\mathrm{int}(c)}\ell_e.
\label{eq:edge_weights}
\end{equation}
Equation~(\ref{eq:zc}) can then be written as
\begin{equation}
z_c=\sum_{e\in\mathrm{int}(c)}
a_e^{(c)}e^{4i\theta_e}.
\end{equation}
Applying Eq.~(\ref{eq:general_participation}) to the edge weights gives
the effective number of independently contributing edges,
\begin{equation}
n_{\rm eff}^{(c)}
\equiv
\frac{1}{
\displaystyle
\sum_{e\in\mathrm{int}(c)}
\left(a_e^{(c)}\right)^2
}
=
\frac{L_c^2}{
\displaystyle
\sum_{e\in\mathrm{int}(c)}\ell_e^2
}.
\label{eq:neff_edges}
\end{equation}
This quantity equals the actual number of interior edges when all edges
have the same length. It is smaller when a few long edges dominate the
phasor sum. Across the dataset the median ratio is
$n_{\rm eff}^{(c)}/N_c\simeq0.73$, where $N_c$ is the number of interior
edges of the domain, so the two measures of domain size differ only by a
roughly constant factor.

For independent and uniformly distributed phases $4\theta_e$,
Eq.~(\ref{eq:weighted_phasor_variance}) gives the exact second moment
\begin{equation}
\left\langle q_c^2\right\rangle_{\rm rand}= \left\langle |z_c|^2\right\rangle_{\rm rand}=
\frac{1}{n_{\rm eff}^{(c)}}.
\label{eq:qc_second_moment}
\end{equation}rather than
When sufficiently many independent edges contribute, the real and
imaginary parts of $z_c$ are approximately Gaussian. Its modulus then
follows the classical Rayleigh distribution, with mean
\begin{equation}
\left\langle q_c\right\rangle_{\rm rand}
\simeq
\frac{\sqrt{\pi}}
{2\sqrt{n_{\rm eff}^{(c)}}}.
\label{eq:qrand}
\end{equation}
The measured values of $q_c$ must therefore be compared with this
finite-size random baseline, rather than with a geometric bound derived
from the clustering tolerance. Note that equation~(\ref{eq:qrand}) relies on the Gaussian approximation, which is
accurate only when many edges contribute, whereas more than half of the
domains contain fewer than ten. The comparison does not depend on this
approximation: the second moment~(\ref{eq:qc_second_moment}) is exact
for any number of edges, and Jensen's inequality gives the
assumption-free bound
\begin{equation}
\left\langle q_c\right\rangle_{\rm rand}
\;\le\;
\sqrt{\left\langle q_c^2\right\rangle_{\rm rand}}
\;=\;
\frac{1}{\sqrt{n_{\rm eff}^{(c)}}},
\label{eq:qrand_bound}
\end{equation}
which exceeds the Rayleigh mean by only $13\%$ ($=1-\sqrt{\pi}/2$). Using this conservative
bound in place of Eq.~(\ref{eq:qrand}) leaves all conclusions below
unchanged.

At the typical effective domain size,
$n_{\rm eff}^{(c)}=42$, Eq.~(\ref{eq:qrand}) gives
$\langle q_c\rangle_{\rm rand}\simeq0.14$, approximately seven times
below the measured mean coherence $0.964$. The comparison is in fact
stronger than this single ratio suggests, because the measured
coherence is essentially independent of domain size whereas the random
baseline is not. The median $q_c$ varies only between $0.969$ and
$0.977$ across five decades of $N_c$
(Fig.~\ref{fig:composite}, bottom), while
Eq.~(\ref{eq:qrand}) decreases as
$\left(n_{\rm eff}^{(c)}\right)^{-1/2}$. The excess over chance
therefore increases from a factor of approximately $2$ in the smallest
domains to approximately $67$ in domains with $N_c>10^3$. Pooling indiscriminately over all $2.84\times10^6$ domains gives a
length-weighted mean ratio of $5.2$, but this understates the effect
because the average is dominated by the very numerous small domains,
for which the finite-size random baseline remains relatively large.
Fewer than $0.1\%$ of domains fall below their own random expectation.
Near-perfect coherence is thus a property of the street fabric, not a
consequence of how the clustering tolerance is chosen. Orientational
grains are internally near-perfect ``crystallites'' everywhere, from
planned grids to organically grown fabrics, and no city is disordered
at the grain scale.

\paragraph{Domain sizes span five decades.}

Let $N_c$ denote the number of interior edges of domain $c$---those
entering Eq.~(\ref{eq:zc})---and $S_c$ its number of nodes. The
distribution of $N_c$ is extremely broad (see Fig.~S6 in Section IV of the Supplementary Material). The median domain contains
only $6$ interior edges and, since $N_c<S_c$ for the median, is a tree:
such domains correspond to cul-de-sacs, branches, and connector stubs,
which are numerous but individually tiny. When domains are weighted by street length, the median rises to
$N_c=74$, with interquartile range $[14,584]$. Street length is spread
almost uniformly across half-decades of $N_c$, so no characteristic
domain size emerges. At the opposite extreme, the largest single
domain in the dataset---in Chicago---contains $137,610$ interior edges
spanning $4285\,\mathrm{km}$. It carries $89\%$ of that city's street
length and has coherence $q_c=0.995$. The urban mosaic therefore ranges
continuously from cities fragmented into many small grains to cities
that are, to a good approximation, single crystals.

\paragraph{Global order is set by mixing.}

Because the step lengths $q_c$ are nearly fixed at a value close to
unity, the length of the phasor walk, and hence the global order
$Q_4^{\mathrm{int}}$, is controlled mainly by the step
\emph{directions}: the number of domains, their weights $w_c$, and
their mutual orientations $\phi_c$. What distinguishes a city such as
Denver (Fig.~\ref{fig:classes}) from a fragmented urban fabric is
therefore not the degree of local order, which is high in both cases,
but the number of internally coherent grains and the way their
orientations are combined. The next subsection makes this statement
quantitative by asking how much city-scale order would remain if the
orientations of the observed grains were completely randomized.

\subsection{A random mosaic null model}
\label{sec:random_mosaic}

A measurable preferred orientation at the scale of the whole city does
not necessarily imply that its different grains are mutually aligned.
Even when grain orientations are completely random, a finite number of
grains generally produces a nonzero resultant, just as a finite random
walk rarely returns exactly to its starting point.

For each city, we therefore construct a random-mosaic null model that
preserves the size and internal order of every observed grain---that is,
its weight $w_c$ and coherence $q_c$---but replaces the phases
$4\phi_c$ by independent random variables uniformly distributed over
$[0,2\pi)$. The grains themselves are thus left unchanged, while all
correlations between their orientations are destroyed. Any order in
excess of the null prediction measures alignment between distinct
grains.

Under this null model, $Z$ is a random walk in the complex plane. Its
mean vector vanishes, while its mean squared length is
\begin{equation}
\left\langle Z\right\rangle_{\rm null}=0,
\qquad
\left\langle |Z|^2\right\rangle_{\rm null}=
\sum_c w_c^2q_c^2.
\label{eq:null_moments}
\end{equation}
The exact scale of the residual city-level order generated by
finite-size fluctuations is therefore
\begin{equation}
\sigma_Z
\equiv
\sqrt{
\left\langle |Z|^2\right\rangle_{\rm null}
}
=
\sqrt{\sum_c w_c^2q_c^2}.
\label{eq:sigma_exact}
\end{equation}

Applying the participation-number definition
Eq.~(\ref{eq:general_participation}) to the grain weights gives the
effective number of grains,
\begin{equation}
N_{\rm eff}
\equiv
\frac{1}{\sum_c w_c^2}.
\label{eq:Neff}
\end{equation}
For $N$ equal-sized grains, $w_c=1/N$ and hence
$N_{\rm eff}=N$. When the grain sizes are heterogeneous,
$N_{\rm eff}$ counts only the grains carrying an appreciable fraction
of the total street length. For example, a city composed of one
dominant domain and many small patches has
$N_{\rm eff}\simeq1$, because $\sum_cw_c^2$ is dominated by the largest
grain. 

Because the grain coherences vary little within a city, we may write
$q_c\simeq\bar q_{\rm city}$. Equation~(\ref{eq:sigma_exact}) then
reduces to
\begin{equation}
\sigma_Z
\simeq
\bar q_{\rm city}
\sqrt{\sum_c w_c^2}=
\frac{\bar q_{\rm city}}{\sqrt{N_{\rm eff}}}.
\label{eq:null}
\end{equation}
Thus, even a completely random mosaic exhibits a residual city-scale
order of magnitude $N_{\rm eff}^{-1/2}$. This residual does not
indicate any collective alignment between grains; it is simply the
finite resultant produced by adding a finite number of randomly
oriented contributions.

When many grains contribute, the real and imaginary parts of $Z$ are
approximately Gaussian by the central limit theorem. The normalized
order
\begin{equation}
R
\equiv
\frac{Q_4^{\mathrm{int}}}{\sigma_Z}
\label{eq:R}
\end{equation}
then follows the Rayleigh distribution
\begin{equation}
P(R)=2R,e^{-R^2},
\qquad
R\geq0.
\label{eq:R_distribution}
\end{equation}
Its mean is
\begin{equation}
\left\langle R\right\rangle_{\rm null}=
\frac{\sqrt{\pi}}{2}
\simeq0.886,
\end{equation}
and its upper-tail probability is
\begin{equation}
P(R>r)=e^{-r^2}.
\label{eq:rayleigh}
\end{equation}
In particular, $R=1$ is not a significance threshold, since
\begin{equation}
P(R>1)=e^{-1}\simeq0.37.
\end{equation}
More than one third of random mosaics therefore have $R>1$. By
contrast,
\begin{equation}
P(R>2)=e^{-4}\simeq1.8\times10^{-2},
\end{equation}
so $R>2$ indicates substantial excess alignment when the Rayleigh
approximation is valid.

For cities with only a few effective grains, finite-size corrections to
the Rayleigh law can be important. We therefore also test the null model
directly by generating
$N_{\mathrm{shuffle}}=5000$ randomized mosaics for every city. If $Q_s$
denotes the order obtained in randomized realization $s$, the Monte
Carlo $p$-value is estimated as
\begin{equation}
p=\frac{
1+
\displaystyle\sum_{s=1}^{N_{\mathrm{shuffle}}}
\mathbf{1}
\left(
Q_s\geq Q_4^{\mathrm{int}}
\right)
}{
N_{\mathrm{shuffle}}+1
}.
\label{eq:pvalue}
\end{equation}
Here $\mathbf{1}(\cdot)$ is the indicator function. The $p$-value is
therefore the probability, under the random-mosaic null model, of
obtaining a city-scale order at least as large as the observed  ($Q_s>Q_4^{\rm int}$). A
small $p$-value means that random grain orientations rarely produce the
measured order.

Across the $1025$ cities, $30\%$ have $p>0.05$: their global order
cannot be distinguished from the finite residual expected for a random
mosaic of their own grains. The remaining $70\%$ have $p<0.05$ and
display stronger alignment between grains than predicted by the null
model. 

\begin{figure}
\centering
\includegraphics[width=0.49\textwidth]{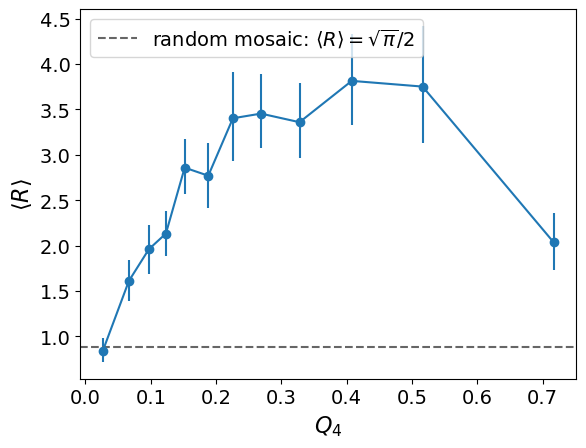}
\caption{
Excess alignment between grains across the structural diagram.
Binned average of
$R=Q_4^{\mathrm{int}}/\sigma_Z$ as a function of $Q_4$
(equal-population bins; error bars show bootstrap $95\%$ confidence
intervals). The dashed line shows the large-$N_{\rm eff}$ Rayleigh
expectation
$\langle R\rangle_{\rm null}=\sqrt{\pi}/2$. Excess alignment is
maximal at intermediate order, where several distinct grains remain
present but share a common orientation.
}
\label{fig:Rmax}
\end{figure}

The strength of this excess alignment is not monotonic in the total
order $Q_4$. As shown in Fig.~\ref{fig:Rmax}, the average ratio $R$ is
largest for cities of intermediate order. The two extremes lie closer
to the random-mosaic prediction for different reasons. Highly
fragmented cities contain many domains, but their orientations are
largely uncorrelated and cancel as in a random mosaic. Near-grid cities, by contrast, are dominated by one or a few large
grains. Their high global order therefore does not result from the
alignment of many distinct domains, but mainly from the fact that most
streets already belong to the same orientational grain.

The strongest departure from randomness therefore occurs between these
two limits. Such cities remain divided into several distinct grains,
but the grains share a common orientational axis. The random-mosaic
comparison thus reveals a mesoscopic form of organization that is not
captured by $Q_4$ alone. The most pronounced coordination between
districts is found not in the most grid-like cities, but in cities where
a polycrystalline structure coexists with a city-wide orientational
constraint.

\section{The morphology diagram}

The quantities $Q_4$ (and $Q_4^{\mathrm{int}}$), $N_{\mathrm{eff}}$, and $R$ describe the
orientational \emph{composition} of a city---how many grains it has and
how they are aligned---but say nothing about the \emph{shape} of the
mosaic. Two cities with the same number of grains may have compact,
well-separated domains or ragged, interpenetrating ones. This geometric
information is exactly what the interface fraction $\gamma$ measures, and
it turns out to be largely independent of the composition. The distinction between the two quantities can be seen directly in the data. Although $Q_4$ and $\gamma$ are anticorrelated, the relation remains weak: a fit $\gamma \sim Q_4^{\beta}$ gives $\beta=-0.178\pm0.009$ and explains only about one quarter of the variance ($R^2=0.258$ in log--log space). Cities with similar global orientational order can therefore still display widely different degrees of fragmentation. On the other hand, $\gamma$ increases only weakly with the number of grains. A fit
$\gamma \sim N_{\mathrm{eff}}^{,\alpha}$ gives
$\alpha = 0.126 \pm 0.010$ ($R^2 = 0.393$ in log--log space), well below
the reference value $\alpha=1/2$ expected if a fixed window were simply
partitioned into $N_{\mathrm{eff}}$ compact grains separated by thin
boundaries. The observed fragmentation is therefore not determined by
the number of grains alone. It also reflects the roughness and
interlocking of their boundaries and thus defines a distinct
morphological axis (Figures in the section III of the Supp. Mat.).

\begin{figure}
	\centering
     \includegraphics[width=0.49\textwidth]{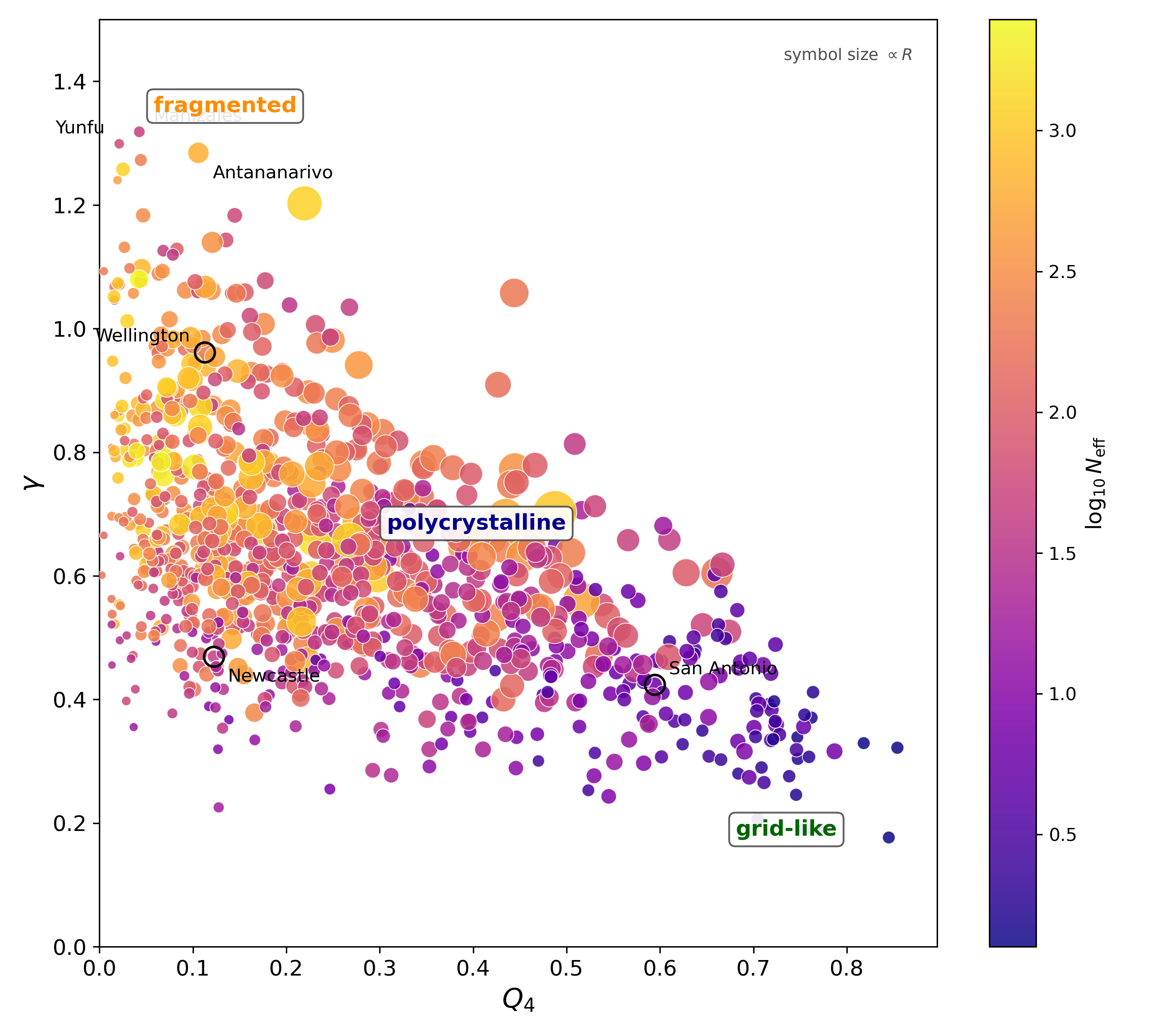}
	\caption{Morphology diagram of urban street networks. Each of the
$1025$ cities is represented by a point in the $(Q_4,\gamma)$ plane
($\delta=10^\circ$; $10$-km analysis discs), where $Q_4$ measures
four-fold orientational order at the scale of the whole city and
$\gamma$ measures the interface fraction of the bearings-community
decomposition. Color encodes $\log_{10}N_{\mathrm{eff}}$, where $N_{\mathrm{eff}}$ is the effective
number of orientational domains, and symbol size represents the
alignment ratio
$R=Q_4^{\mathrm{int}}/\sigma_Z$
[Eq.~(\ref{eq:R})], which compares the internal orientational order of
each city with that expected for a random mosaic having the same domain
weights and coherences. The distribution is continuous, and the labels indicate three limiting regimes:
grid-like cities, dominated by a very small number of orthogonal frames;
polycrystalline cities, composed of several distinct and internally
ordered domains; and fragmented cities, containing many small,
strongly interlocked domains. The decrease of $N_{\mathrm{eff}}$ toward
the grid-like region is consistent with the grain-counting mechanism of
Eq.~(\ref{eq:null}): internal city-scale order generally increases as
the effective number of independently oriented domains decreases
[$\mathrm{corr}(\log Q_4^{\mathrm{int}},
\log N_{\mathrm{eff}})=-0.57$], while the mean domain-level coherence
remains nearly constant ($\langle\bar q_{\rm city}\rangle=0.964\pm0.014$). Ringed symbols mark a representative city from each regime, and the
other labeled cities indicate
notable outliers in the diagram.}
	\label{fig:clustering}
\end{figure}

These two axes together define the \emph{morphology diagram} of
Fig.~\ref{fig:clustering}, in which each of the $1025$ cities is a point
in the $(Q_4,\gamma)$ plane, colored by its number of grains
$N_{\mathrm{eff}}$ and sized by its alignment ratio $R$. Four features
summarize what the diagram shows.

\emph{(i) The distribution is continuous.} Cities do not fall into
separate clusters; they fill the plane, and we therefore draw no
dividing lines. What organizes the diagram is instead a set of three
\emph{limiting regimes} at its corners, in the same spirit as an
order-parameter diagram in condensed matter. \emph{Grid-like} cities
(large $Q_4$, small $\gamma$) are dominated by a single orthogonal
frame---the analogue of a single crystal. \emph{Polycrystalline} cities
(small to moderate $Q_4$, small $\gamma$) are made of several large,
internally ordered domains with different orientations, meeting along
sharp boundaries. \emph{Fragmented} cities (small $Q_4$, large $\gamma$)
consist of many small, strongly interlocked domains. Most cities lie
somewhere between these limits rather than at them.

\emph{(ii) The colour reveals the mechanism.} The number of grains
$N_{\mathrm{eff}}$ varies smoothly across the plane, growing from the
grid-like corner (few grains) toward the fragmented one (many). This is
the null model~(\ref{eq:null}) at work: because grains are internally
ordered in every city ($\langle \bar q_{\rm city}\rangle \simeq 0.96$, nearly constant), a city's
global order is set mainly by how many grains it averages over---the
more grains, the more their orientations cancel, and the smaller $Q_4$.
Moving across the diagram is therefore, in essence, \emph{grain
coarsening}: the same distinction between a few large grains and many
small ones that separates a single crystal from a fine-grained solid.

\emph{(iii) The vertical axis adds what the color cannot.} At a fixed
color---fixed $N_{\mathrm{eff}}$---cities still span a broad range of
$\gamma$. Two cities can be built from the same number of equally ordered
grains and yet differ in how those grains interlock, one tiled from
compact blocks, the other shot through with ragged boundaries.

\emph{(iv) Alignment peaks in the middle.} The symbol size proportional to the ratio $R$ is not uniform: its binned average $\langle R\rangle$ is
largest at intermediate $Q_4$ and drops to the random-mosaic level at
both ends (Fig.~\ref{fig:Rmax}). The reason was given above: writing
$R = (Q_4^{\mathrm{int}}/\bar q_{\rm city})\sqrt{N_{\mathrm{eff}}}$, the near-grid
corner has too few grains for any alignment to register
($R \le \sqrt{N_{\mathrm{eff}}}$), while the fragmented corner is
statistically indistinguishable from randomly oriented grains. Only in
between do cities combine many grains \emph{with} a shared orientation,
making them the least random fabrics in the dataset.

\begin{figure*}
    \centering
    \includegraphics[width=0.95\linewidth]{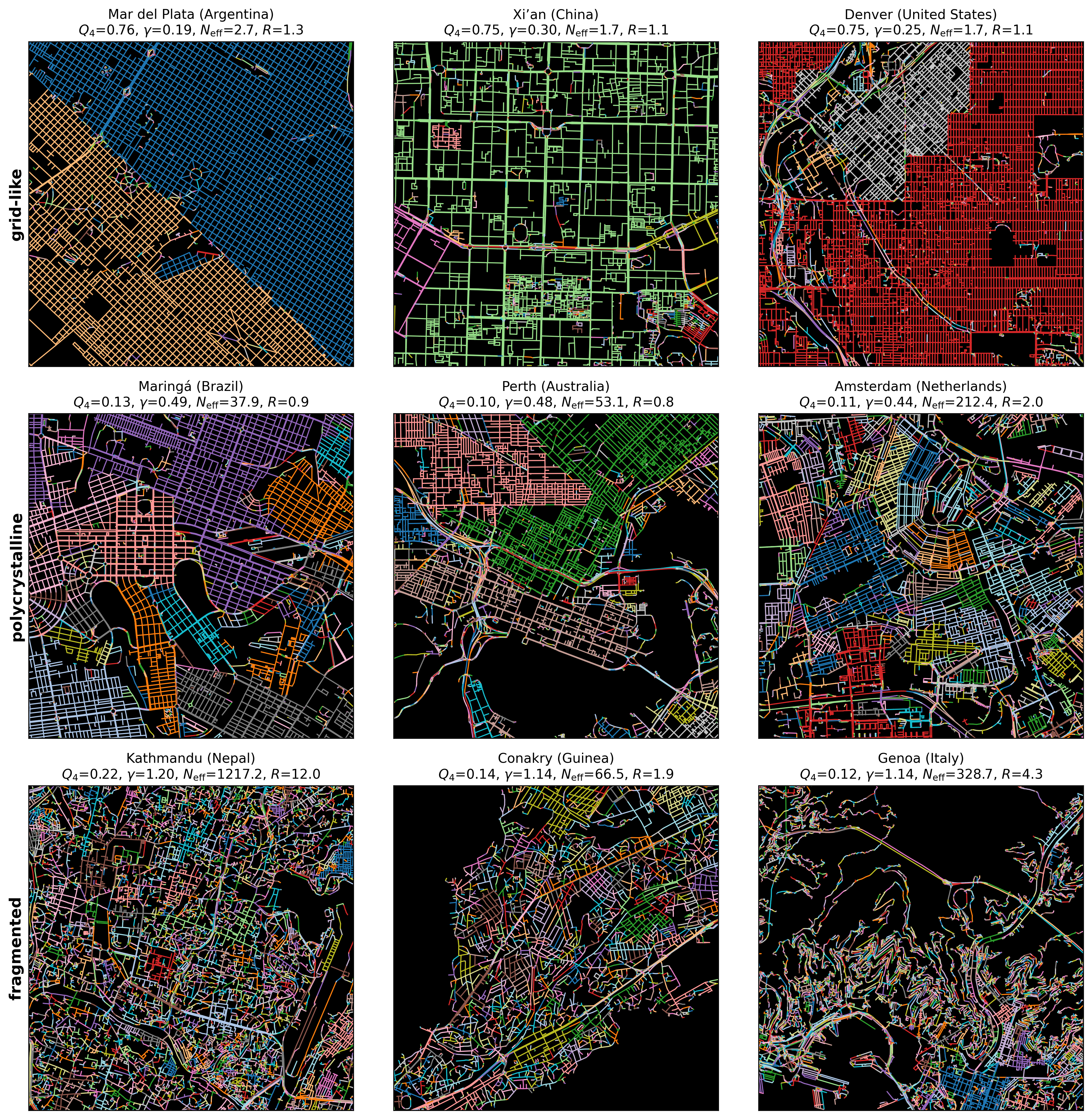}
    \caption{Decomposition of street networks into bearings
communities for nine representative cities ($\delta = 10^\circ$) where edges are colored by
community. For the sake of visibility, we show here the central $6\,\mathrm{km}\times6\,\mathrm{km}$ region of the $10$-km
analysis disc. The three rows illustrate the three limiting
regimes of urban morphology identified in this work
(Fig.~\ref{fig:clustering}). Top row, \emph{grid-like}: Mar del
Plata (Argentina), Xi'an (China), and Denver (United States), each
dominated by very few orthogonal frames extending across the urban
area. Middle row,
\emph{polycrystalline}: Maringá (Brazil), Perth (Australia), and
Amsterdam (Netherlands), composed of several large, internally
ordered domains with distinct orientations meeting along sharp
boundaries. Bottom row, \emph{fragmented}: Kathmandu (Nepal),
Conakry (Guinea), and Genoa (Italy), where numerous small domains
interlock and interfaces pervade the urban fabric. Each regime is
represented by cities from several continents, indicating that
these morphologies reflect general growth processes rather than
region-specific styles. Panel titles report the quantitative
measures: the global orientational order $Q_4$
[Eq.~(\ref{eq:Q4})], the interface fraction $\gamma$, the effective
number of domains $N_{\mathrm{eff}}$ [Eq.~(\ref{eq:Neff})], and the
alignment ratio $R$ [Eq.~(\ref{eq:R})]. The values show a general
increase in $N_{\mathrm{eff}}$ and $\gamma$ from the grid-like to
the fragmented regime. Street-network
data are from OpenStreetMap~\cite{OSM}.}
\label{fig:classes}
\end{figure*}

Representative cities for each limiting regime are shown in Fig.~\ref{fig:classes}.

\section{Discussion}

By decomposing each city into orientational grains, our framework turns the qualitative typology of urban form into a quantitative,
physically grounded description of its mesoscopic structure. In particular, this framework resolves ambiguities that global angular statistics cannot capture.

The central result is a separation between local and city-wide order.
At the grain scale, bearings communities are highly coherent in almost
all cities, so cities differ little in the quality of their local order.
What varies is the organization of the mosaic: how many grains are
present, how large they are, and how their orientations are related.

The random-mosaic comparison shows that part of the global order is a
finite-grain effect: cities made of only a few large grains remain
ordered even when those grains are randomly oriented. This accounts for
the observed order in about \(30\%\) of cities. In the remaining
\(70\%\), the grains are more aligned than expected by chance, revealing
orientational coordination across distinct parts of the city.

Urban fabrics therefore differ not because some cities have locally
better grids than others, but because their coherent grains are assembled
in different ways. The relevant distinction is between cities dominated
by a few grains, random polycrystalline mosaics, and mosaics whose grains
share a common large-scale orientation. In other words, much of the
apparent diversity of urban orientational order is, at bottom, a grain-counting effect: a city looks ordered mainly because it is built from few large grains, not because its fabric is locally more crystalline. In this picture the elementary building blocks of urban form are the mesoscopic grains themselves, and cities interpolate continuously between the single-crystal, polycrystalline, and fragmented limits according to how many grains they contain and
how those grains are mixed.

The interface fraction \(\gamma\) provides complementary information
about the geometry of the mosaic. It is only weakly related to the
effective number of grains, showing that fragmentation is not determined
by grain count alone. Cities with similar numbers of grains can differ
strongly in whether these grains form compact domains or irregular,
interlocking patches. Moreover, interface streets partly retain the
orientations of the neighbouring grains, indicating that grain
boundaries are structured components of the urban fabric rather than
orientationally random separators.

Taken together, these regimes suggest different pathways of urban
formation (Fig.~\ref{fig:classes}). Grid-like cities with few effective
grains are consistent with large-scale coordinated development.
Polycrystalline fabrics may result from the juxtaposition or coalescence
of districts developed at different times or under different planning
rules. Highly fragmented fabrics, characterized by many interlocking
grains and a large interface fraction, may instead reflect incremental
growth, physical constraints, or the disruption of existing grids by
major infrastructure. These interpretations remain hypotheses that
should be tested against the historical development of individual
cities.

The polycrystalline analogy is therefore quantitative rather than purely
descriptive. Bearings communities define the grains, their orientations
determine the global order, and the interfaces identify where one local
street organization gives way to another. In urban terms, these
boundaries may mark changes in planning principles, development periods,
or infrastructure constraints, providing a quantitative counterpart to
Lynch's distinction between districts and edges~\cite{lynch1964image}.
A natural next step is to study how grains appear, expand, merge, and
reorient through time, and thereby connect the present morphology of a
street network to the historical processes that produced it.

\bibliography{bibfile_street_network}
\bibliographystyle{ieeetr}

\end{document}


\title{Supplemental Material\\ Geometry of Urban Order}

\author{Marc Barthelemy}
\email{marc.barthelemy@ipht.fr}
\affiliation{Universit\'e Paris-Saclay, CNRS, CEA, Institut de Physique Théorique, 91191 Gif-sur-Yvette, France}
\affiliation{Centre d'Analyse et de Math\'ematique Sociales (CNRS/EHESS), Paris, France}
\affiliation{Complexity Science Hub, Vienna, Austria}

\author{Geoff Boeing}
\email{boeing@usc.edu}
\affiliation{Department of Urban Planning and Spatial Analysis,
Sol Price School of Public Policy, University of Southern California,
301A Lewis Hall, Los Angeles, CA 90089-0626, USA}

\date{\today}

\maketitle

\section{Data, preprocessing, and the dominant node orientation}
\label{sec:SI_phi}

This section describes the dataset and preprocessing, details the
definition of the dominant orientation assigned to each node, clarifies
its algorithmic implementation, and illustrates its interpretation
through simple representative examples.

\subsection{Data and preprocessing}
\label{sec:SI_data}

Street-network data were obtained from OpenStreetMap using country-level
extracts downloaded from Geofabrik. The extracts were processed locally
with osmium-tool~1.16.0 and pyrosm~0.7.0; OSMnx~2.0.7 was used for
coordinate projection.

The city list was constructed from the GeoNames database of populated
places. We first retained cities with a reported population of at least
$10^5$ inhabitants and then selected up to the $200$ most populous
cities on each continent. City centers were defined directly from the
latitude and longitude coordinates provided by GeoNames; no geocoding
service was used. After removing duplicate city--country entries and
retaining only cities for which the complete network and community
analysis could be performed, the final dataset contained $1025$ cities.

For each city, we extracted the drivable street network inside a disc of
radius $10$~km centered on the GeoNames coordinates. The same observation
window was used for all cities because several quantities considered
here, including the interface measure $\gamma$ and the effective number
of domains $N_{\mathrm{eff}}$, depend on the size of the analyzed region.
A fixed window is therefore necessary for meaningful comparisons across
cities.

The network was projected onto a local metric coordinate system and
restricted to nodes lying within the $10$-km disc. It was then converted
to an undirected simple graph, with self-loops removed. Each retained
node is specified by its projected coordinates $(x_n,y_n)$, and each
edge $e$ by its metric length $\ell_e$. For the orientational analysis,
the bearing $\theta_e$ of each edge is computed from the coordinates of
its endpoints and defined modulo $\pi$, since a street segment has no
intrinsic direction.

Bearings communities containing fewer than $5$ nodes are excluded from
the community-level statistics. Two additional choices
concern the definition of the local orientation. At a node $n$, during the computation of the dominant orientation $\phi_n$, we
compute the fourth-harmonic resultant
\begin{equation}
R_n =
\left|
\left\langle e^{4i\theta_{nj}}\right\rangle_{j\sim n}
\right|,
\end{equation}
where the average is taken over the edges incident to the node. When
$R_n$ is very small, no dominant local orientation can be assigned. This
occurs, for example, at an exactly symmetric three-fold junction. Nodes
with $R_n<R_{\min}$, with $R_{\min}=0.1$, are therefore treated as
singletons: they can neither initiate nor join a bearings community.

All results reported in the main text use an angular tolerance
$\delta=10^\circ$. 

\subsection{Definition of the dominant node orientation}

The dominant orientation at node $n$ is defined from the set of bearings of all incident
edges. To treat orthogonal directions on equal footing---reflecting the fourfold symmetry
of grid-like patterns---we compute the circular mean of the bearings on the
$[0,\pi/2)$ interval:
\begin{equation}
\phi_n
\;=\;
\frac{1}{4}\,\arg\!\left\langle\,
e^{\,4\mathrm{i}\,\theta_{nj}}
\,\right\rangle_{j\sim n}
\;\bmod\;\tfrac{\pi}{2}\,,
\qquad
\phi_n\in\bigl[0,\tfrac{\pi}{2}\bigr).
\label{eq:phi_def}
\end{equation}
Here $j\sim n$ indicates that $j$ and $n$ are neighbouring nodes, and the angle brackets
denote the arithmetic mean over all incident edges. The factor of~$4$ in the exponent maps
the fourfold-symmetric orientations onto the full circle before averaging, ensuring that
angles near opposite ends of the $[0,\pi/2)$ interval are correctly recognised as nearly
equivalent. The $\bmod\;\pi/2$ operation maps the result of $\frac{1}{4}\arg(\cdot)$,
which lies in $(-\pi/4,\,\pi/4]$, back into $[0,\pi/2)$.

To compare orientations consistently on this compact interval, angular differences are
measured using the distance
\begin{equation}
d_{\pi/2}(\alpha,\beta)
=
\min\!\left(
|\alpha-\beta|,\;
\tfrac{\pi}{2}-|\alpha-\beta|
\right),
\label{eq:dpi2}
\end{equation}
which corresponds to the shortest arc length on a circle of circumference $\pi/2$.

\subsection{Why a circular mean is necessary}
\label{sec:SI_circular}

A natural first choice for a robust central tendency would be the linear median of the reduced bearings $\{\theta_{nj}\bmod\pi/2\}$. However, this fails for orientations near the wrapping boundary of the $[0,\pi/2)$ interval, because the linear median does not respect the periodic topology.

\paragraph{Failure of the linear median.}
Consider a four-way intersection on a nearly cardinal grid, with bearings
$\{1^\circ, 89^\circ, 181^\circ, 271^\circ\}$.
After reduction modulo $90^\circ$, the set becomes $\{1^\circ, 89^\circ, 1^\circ, 1^\circ\}$, whose median is $1^\circ$, and this leads to a correct result in this case. Now consider a neighbouring
node with slightly different geometry, carrying bearings
$\{89^\circ, 179^\circ, 1^\circ, 271^\circ\}$.
After reduction modulo $90^\circ$: $\{89^\circ, 89^\circ, 1^\circ, 1^\circ\}$.
The sorted values are $\{1, 1, 89, 89\}$, and the median is
$(1+89)/2=45^\circ$---a completely spurious value, since the node sits on the same
cardinal grid as its neighbour. The underlying cause is that the values $1^\circ$ and
$89^\circ$ are close on the circle ($d_{\pi/2}=2^\circ$) but far apart on the real line
($88^\circ$), and the linear median treats them as distant.

For example, in a city like Minneapolis, whose main grid runs almost exactly along the cardinal
directions, the node orientations cluster near~$0^\circ$ (equivalently~$90^\circ$),
placing them directly on top of the wrapping boundary. The median artefact therefore
affects a significant fraction of nodes and produces small spurious ``island'' clusters
within otherwise uniform grid neighbourhoods.

\paragraph{Circular mean resolution.}
The circular mean in Eq.~\eqref{eq:phi_def} avoids this problem entirely. For the same
bearings $\{89^\circ, 179^\circ, 1^\circ, 271^\circ\}$, the exponentiation
$e^{4\mathrm{i}\theta}$ maps each bearing onto the unit circle: the values $1^\circ$ and
$89^\circ$ (modulo $90^\circ$) both map to unit vectors near angle~$0$ (equivalently
$2\pi$), so their complex average points unambiguously towards~$0$, yielding
$\phi_n\approx 0^\circ$---the correct grid orientation. More generally, the circular mean
treats the interval $[0,\pi/2)$ as a true periodic domain with no privileged boundary,
eliminating wrapping artefacts regardless of the grid orientation.

\subsection{Use in the clustering algorithm}

Bearings communities are constructed by grouping spatially connected nodes whose dominant
orientations are compatible. Starting from an unassigned seed node~$s$, graph-neighbours
are iteratively added provided their dominant orientation satisfies
\begin{equation}
d_{\pi/2}(\phi_n,\phi_s) < \delta,
\label{eq:coherence}
\end{equation}
where $\delta$ is a fixed angular tolerance. The process continues until no further
compatible neighbours can be included. Comparing every candidate to the \emph{seed}
orientation~$\phi_s$ rather than to its immediate predecessor constitutes a \emph{strict}
criterion that prevents gradual orientational drift within a single cluster and ensures
that each bearings community corresponds to a well-defined orientational domain.
Nodes with undefined orientation ($R_n < R_{\min}$; see
Sec.~\ref{sec:SI_data}) are treated as singletons: they can neither
seed a community nor be absorbed into one.

\subsection{Illustrative examples}

\paragraph{Orthogonal grid intersection.}
Consider a four-way intersection with incident bearings
$\{0^\circ,90^\circ,180^\circ,270^\circ\}$.
The complex mean is
$\frac{1}{4}(e^{0}+e^{4\mathrm{i}\cdot\pi/2}+e^{4\mathrm{i}\cdot\pi}
+e^{4\mathrm{i}\cdot 3\pi/2})
= \frac{1}{4}(1+1+1+1) = 1$,
so $\phi_n = \frac{1}{4}\arg(1) = 0^\circ$.
Although the intersection connects two perpendicular streets, it is
assigned a single orientational class, consistent with fourfold symmetry.

\paragraph{T-junction.}
For a T-shaped intersection with bearings
$\{0^\circ,180^\circ,90^\circ\}$, the complex mean is
$\frac{1}{3}(e^{0}+e^{4\mathrm{i}\cdot\pi}+e^{4\mathrm{i}\cdot\pi/2})
= \frac{1}{3}(1+1+1)=1$,
yielding $\phi_n=0^\circ$. The dominant orientation reflects the axis of the main road.

\paragraph{Rotated grid.}
For a locally rotated grid with bearings
$\{15^\circ,105^\circ,195^\circ,285^\circ\}$, all four contributions
$e^{4\mathrm{i}\theta}$ have the same phase $4\times 15^\circ=60^\circ$, giving
$\phi_n = 60^\circ/4 = 15^\circ$. The dominant orientation captures the local grid
rotation independently of the global reference frame.

\paragraph{Irregular intersection with a side street.}
Consider bearings
$\{10^\circ,100^\circ,190^\circ,280^\circ,55^\circ\}$.
The first four edges contribute unit vectors at phase $4\times10^\circ=40^\circ$, while
the oblique street contributes a vector at $4\times55^\circ=220^\circ$, nearly
antiparallel. The complex mean is dominated by the four coherent contributions, yielding
$\phi_n\approx 10^\circ$. The oblique side street has little influence, demonstrating the
robustness of the circular mean to outliers when the majority of edges are coherent.

\paragraph{Symmetric three-fold junction (undefined orientation).}
For a perfectly symmetric three-fold junction with bearings
$\{0^\circ, 120^\circ, 240^\circ\}$, the phases after the
$e^{4\mathrm{i}\theta}$ mapping are $0^\circ$, $120^\circ$, and
$240^\circ$ (modulo $360^\circ$), three unit vectors summing to zero.
The resultant $R_n$ vanishes, $\arg(0)$ is undefined, and no meaningful
dominant orientation exists: such nodes are excluded from the
orientational analysis, as described in Sec.~\ref{sec:SI_data}.

\paragraph{Near-boundary orientation (the wrapping case).}
Consider bearings $\{2^\circ, 88^\circ, 182^\circ, 272^\circ\}$. After the
$e^{4\mathrm{i}\theta}$ mapping, the phases are $8^\circ$, $352^\circ$, $8^\circ$, and
$8^\circ$---all clustered near~$0^\circ$ on the unit circle. The circular mean gives
$\phi_n\approx 1^\circ$, correctly identifying the near-cardinal orientation. A linear
median of the reduced bearings $\{2, 88, 2, 2\}$ would also give $2^\circ$ in this case,
but replacing one bearing by $92^\circ$ (reduced to $2^\circ$) versus $88^\circ$ can tip
the median to $45^\circ$, as discussed in Sec.~\ref{sec:SI_circular}.

\subsection{Physical interpretation}

The dominant orientation $\phi_n$ should be interpreted as the locally preferred street
axis at an intersection. By assigning a single orientational degree of freedom to each
node, the network can be viewed as a field of local orientations $\{\phi_n\}$. Bearings
communities then emerge as contiguous regions where this field is approximately uniform,
analogous to orientational grains in polycrystalline materials, separated by interfaces
where the preferred axis changes abruptly.

\clearpage
\newpage

\section{Cities with extreme mean misorientation}
\label{sec:SI_misorientation_extremes}

To illustrate the meaning of the city-level mean misorientation
$\langle\Delta\phi\rangle$, we identify the cities with the smallest
and largest values among networks containing more than $10^{4}$ nodes.
The threshold avoids selecting small networks for which the mean may be
dominated by a limited number of interfaces.

\begin{figure*}[b]
\centering
\includegraphics[width=\textwidth]
{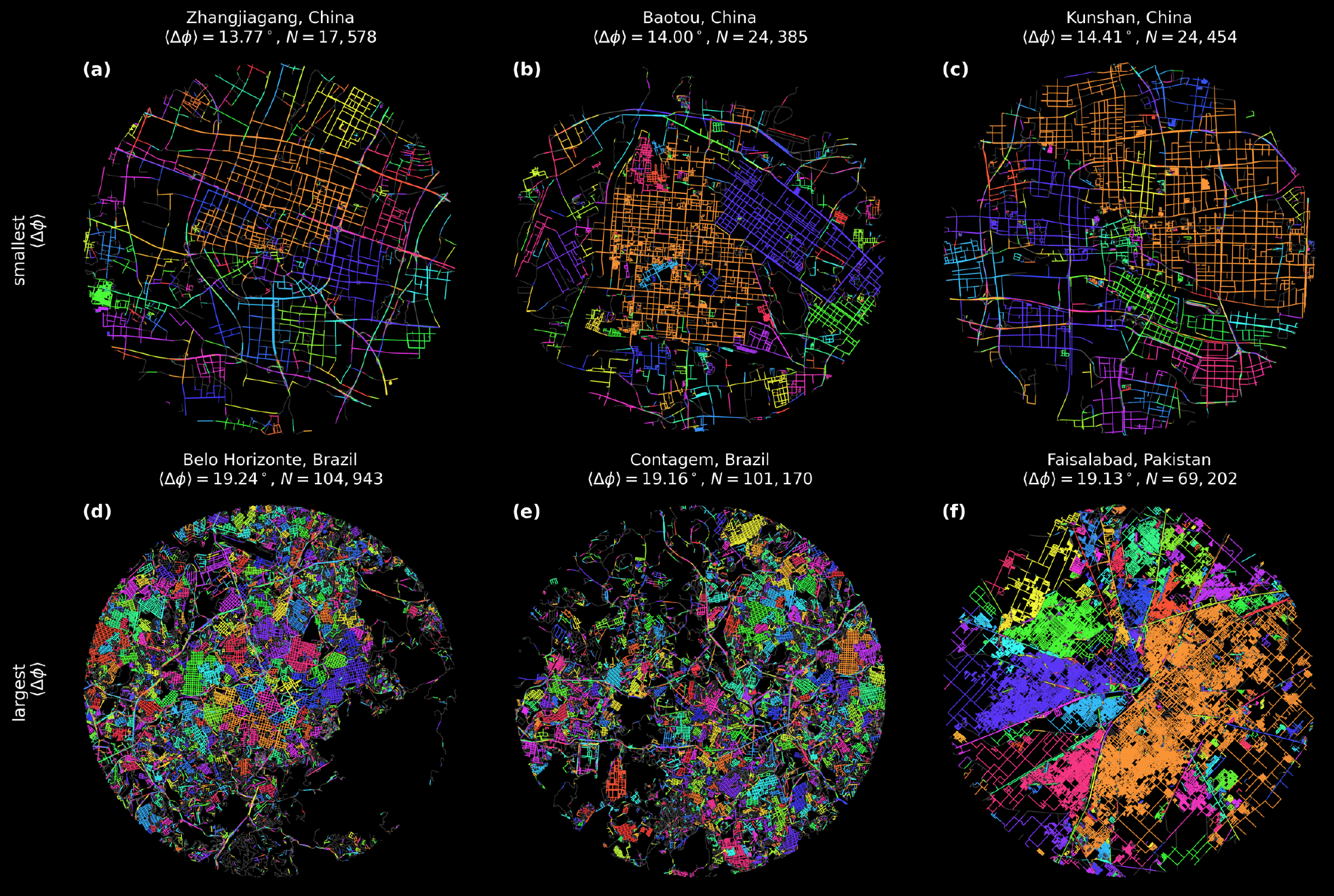}
\caption{Street-network maps of the cities with the three smallest
(top row) and three largest (bottom row) length-weighted mean
misorientations $\langle\Delta\phi\rangle$, after restricting the
comparison to networks containing more than $10^{4}$ nodes.
From left to right, the top row shows Zhangjiagang
($13.77^\circ$; $17,578$ nodes), Baotou
($14.00^\circ$; $24,385$ nodes), and Kunshan
($14.41^\circ$; $24,454$ nodes). The bottom row shows Belo
Horizonte ($19.24^\circ$; $104,943$ nodes), Contagem
($19.16^\circ$; $101,170$ nodes), and Faisalabad
($19.13^\circ$; $69,202$ nodes). Colors identify retained
bearings communities, while gray edges correspond to interfaces,
communities below the minimum-size threshold, or unassigned edges.
Small values of $\langle\Delta\phi\rangle$ indicate that adjacent
communities tend to have similar preferred orientations, whereas
large values indicate stronger orientational contrasts across
community boundaries.}
\label{fig:misorientation_extremes}
\end{figure*}

The three smallest values are found for Zhangjiagang
($\langle\Delta\phi\rangle=13.77^\circ$), Baotou
($14.00^\circ$), and Kunshan ($14.41^\circ$), all in China
(Fig.~\ref{fig:misorientation_extremes}, top row). In these cities,
neighboring bearings communities tend to have relatively similar
orientations. Their street fabrics are divided into many distinct
domains, but the change in orientation across their interfaces is, on
average, comparatively small.

The largest values are observed for Belo Horizonte
($19.24^\circ$), Contagem ($19.16^\circ$), and Faisalabad
($19.13^\circ$) (Fig.~\ref{fig:misorientation_extremes}, bottom row).
Here, adjacent communities tend to meet at larger angular differences,
indicating stronger orientational contrasts between neighboring
districts. These examples show that the mean misorientation measures a
property distinct from the number or size of the communities: it
quantifies how sharply their preferred street orientations differ
across interfaces.

\clearpage
\newpage
\section{Interface analysis}
\label{sec:SI_interface}

\subsection{Relation between the interface measures $\Gamma$, $f_\partial$, and $\gamma$}

Interfaces between bearings communities quantify the spatial
fragmentation of a city's orientational fabric. Let $I$ denote the total
length of street segments lying between distinct bearings communities,
with each inter-community segment counted once, and let $L$ be the total
street length of the city. The global interface fraction is defined as
\begin{equation}
\Gamma = \frac{I}{L}.
\label{eq:Gamma_definition}
\end{equation}
This is the quantity denoted $f_\partial$ in the main text. It gives the
fraction of the street network occupied by boundaries between
orientational domains.

At the community level, let $\Lambda_c$ be the total interface length
incident on community $c$, and let $L_c$ be its internal street length,
excluding inter-community segments. Because each interface separates two
communities, it is assigned to both of them when the community-level
interface lengths are summed. Consequently,
\begin{equation}
\sum_c \Lambda_c = 2I,
\qquad
\sum_c L_c = L-I.
\label{eq:interface_sums}
\end{equation}
The interface-to-bulk ratio used in the main text is therefore
\begin{equation}
\gamma
=
\frac{\sum_c \Lambda_c}{\sum_c L_c}
=
\frac{2I}{L-I}
=
\frac{2\Gamma}{1-\Gamma}.
\label{eq:gamma_Gamma}
\end{equation}
Equivalently,
\begin{equation}
\Gamma = \frac{\gamma}{\gamma+2}.
\label{eq:Gamma_gamma}
\end{equation}

Thus, $f_\partial=\Gamma$ and $\gamma$ contain the same information but
use different normalizations. The fraction $\Gamma$ is bounded between
zero and one and measures the proportion of the network lying on
interfaces. By contrast, $\gamma$ compares the double-counted community
boundary length with the internal street length and is naturally
interpreted as an effective surface-to-volume ratio of the bearings
communities. Small values correspond to large, internally coherent
domains separated by relatively few boundary streets, whereas large
values indicate smaller, more irregular, or more strongly intermixed
domains.

The final dataset contains $1025$ cities with exact values of
$Q_4^{\mathrm{int}}$. The values of $\gamma$ were recomputed from the
grain-level internal and interface lengths for all cities, except for one
case in which Eq.~\eqref{eq:gamma_Gamma} was used as a fallback. Across
the dataset,
\begin{equation}
0.177 \leq \gamma \leq 3.042,
\end{equation}
corresponding through Eq.~\eqref{eq:Gamma_gamma} to
\begin{equation}
0.081 \leq \Gamma \leq 0.603.
\end{equation}

As a consistency check, the directly computed values of $\gamma$ were
compared with $2\Gamma/(1-\Gamma)$. The median absolute difference was
\begin{equation}
9.992\times 10^{-15},
\end{equation}
and the maximum absolute difference was
\begin{equation}
2.646\times 10^{-13}.
\end{equation}
These discrepancies are at the level of floating-point precision,
confirming that the interface-counting convention and the definitions of
$\Gamma$ and $\gamma$ are implemented consistently throughout the
analysis.

\subsection{Relation between $Q_4$ and $Q_4^{\mathrm{int}}$}
\label{sec:SI_q4int}

The full-network order parameter $Q_4$ includes all street segments,
whereas $Q_4^{\mathrm{int}}$ is computed only from edges lying within
bearings communities. Their difference therefore measures the effect of
inter-community interface edges, which are not assigned to any single
community.

To make this relation explicit, let the complex fourth-order parameters
of the interior and interface edges be denoted by
\begin{equation}
Z_4^{\mathrm{int}}
=
\frac{1}{L-I}
\sum_{e\in\mathrm{int}}
\ell_e,e^{4i\phi_e},
\qquad
Z_4^{\partial}
=
\frac{1}{I}
\sum_{e\in\partial}
\ell_e,e^{4i\phi_e},
\end{equation}
where $\ell_e$ and $\phi_e$ are the length and bearing of edge $e$,
respectively. Since the interior and interface edges occupy fractions
$1-\Gamma$ and $\Gamma$ of the total street length, the full-network
complex order parameter satisfies
\begin{equation}
Z_4
=
(1-\Gamma)Z_4^{\mathrm{int}}
+
\Gamma Z_4^{\partial}.
\label{eq:Q4_complex_decomposition}
\end{equation}
The corresponding scalar order parameters are
\begin{equation}
Q_4 = \left|Z_4\right|,
\qquad
Q_4^{\mathrm{int}} = \left|Z_4^{\mathrm{int}}\right|.
\end{equation}
Because Eq.~\eqref{eq:Q4_complex_decomposition} is a vector sum in the
complex plane, there is in general no exact scalar relation between
$Q_4$ and $Q_4^{\mathrm{int}}$.

If interface-edge orientations were completely random, their complex
contribution would approximately cancel,
$Z_4^{\partial}\simeq 0$, yielding
\begin{equation}
Q_4 \simeq (1-\Gamma)Q_4^{\mathrm{int}}.
\label{eq:Q4_random_interfaces}
\end{equation}
The interface edges therefore generally reduce the measured global
order, although the magnitude of this reduction depends on both their
length fraction and their orientational coherence relative to the
interior streets.

Across the $1025$ cities, $Q_4$ and $Q_4^{\mathrm{int}}$ are very
strongly proportional. A least-squares fit constrained to pass through
the origin gives
\begin{equation}
Q_4 = a,Q_4^{\mathrm{int}},
\qquad
a = 0.844 \pm 0.002,
\label{eq:Q4_Q4int_fit}
\end{equation}
where the uncertainty is the standard error of the fitted slope. The
corresponding $95\%$ confidence interval is
\begin{equation}
0.840 \leq a \leq 0.848,
\end{equation}
and the uncentred coefficient of determination is
\begin{equation}
R^2_{\mathrm{unc}} = 0.995.
\end{equation}
Thus, including interface edges reduces the order parameter by roughly
$16\%$ on average at fixed $Q_4^{\mathrm{int}}$, while preserving almost
all of the cross-city variation.

The city-level relative interface correction,
\begin{equation}
\Delta_4
=
\frac{\left|Q_4-Q_4^{\mathrm{int}}\right|}
{Q_4^{\mathrm{int}}},
\label{eq:Q4_interface_correction}
\end{equation}
has a median value of $17.9\%$ and a mean value of $19.8\%$. The
difference between these values and $1-a=15.6\%$ reflects the
city-to-city variation around the global through-origin fit. Overall,
the close proportionality shows that removing interface edges changes
mainly the amplitude of the measured orientational order, rather than
the relative ordering of cities.

Interface abundance is only weakly related to interior orientational
order. A power-law fit gives
\begin{equation}
\gamma
=
0.476,
\left(Q_4^{\mathrm{int}}\right)^{-0.160},
\label{eq:gamma_Q4int_fit}
\end{equation}
with
\begin{equation}
R^2 = 0.190.
\end{equation}
The negative exponent indicates that cities with stronger interior
orientational order tend, on average, to have fewer interfaces relative
to their internal street length. However, the low value of $R^2$ shows
that this relation explains only a small fraction of the cross-city
variation: the orientational coherence within grains and the amount of
interface between grains remain largely distinct aspects of urban
street-network organization.

\begin{figure}[h!]
\centering
\includegraphics[width=0.55\textwidth]
{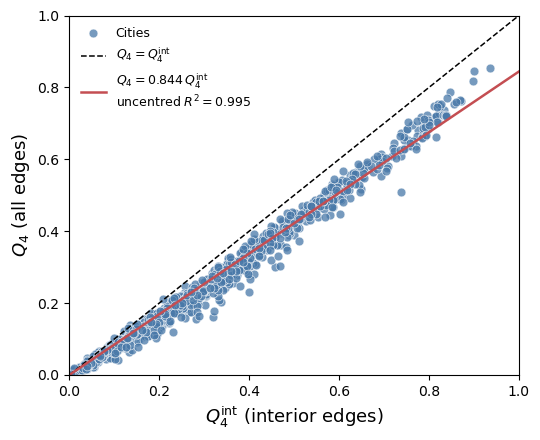}
\caption{Relation between the full-network order parameter $Q_4$
and the interior-edge order parameter $Q_4^{\mathrm{int}}$ for the
$1025$ cities. The solid line is the through-origin fit
$Q_4=0.844,Q_4^{\mathrm{int}}$. The dashed line shows the
prediction $Q_4=(1-\langle\Gamma\rangle)Q_4^{\mathrm{int}}$
obtained when interface-edge orientations are assumed to be
random. The close proportionality indicates that interface edges
primarily attenuate the amplitude of the global order parameter
without substantially changing the relative ordering of cities.}
\label{fig:q4fit}
\end{figure}

\subsection{Distribution of the interface ratio $\gamma$}

\begin{figure}
\centering
\includegraphics[width=0.60\textwidth]
{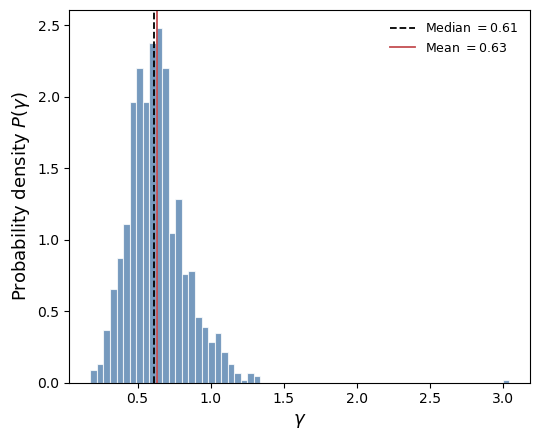}
\caption{Normalized probability distribution $P(\gamma)$ of the
interface-to-bulk ratio $\gamma$ across the $1025$ cities. The
distribution has mean $0.631$, standard deviation $0.203$, and
median $0.611$; its interquartile range is
$[0.499,0.728]$.}
\label{fig:Pgamma}
\end{figure}

We next examine the distribution of the interface-to-bulk ratio
$\gamma$ across cities. Figure~\ref{fig:Pgamma} shows the normalized
probability density $P(\gamma)$ for the full dataset of $1025$ cities.
The distribution has mean and standard deviation
\begin{equation}
\langle \gamma \rangle = 0.631 \pm 0.203,
\end{equation}
and a median value of $0.611$. Its interquartile range is
\begin{equation}
0.499 \leq \gamma \leq 0.728,
\end{equation}
while the full range extends from $0.177$ to $3.042$.

The mean lies slightly above the median, and the upper tail extends far
beyond the interquartile range, indicating a moderately right-skewed
distribution with a small number of particularly fragmented cities.
Small values of $\gamma$ correspond to orientational domains with long
internal street networks relative to their interfaces. Large values
instead indicate that a substantial fraction of community-level street
length lies along boundaries, as occurs when the urban fabric is
partitioned into many small, irregular, or strongly interlocked
bearings communities.

Cities dominated by large and
spatially coherent communities tend to occupy the low-$\gamma$ regime,
whereas cities containing many small or irregular communities tend to
have larger values. Nevertheless, $\gamma$ is not determined by
community size alone: it also depends on boundary roughness, spatial
interlocking, and the distribution of street length within the
communities. It thus provides a mesoscopic measure of fragmentation
that is complementary to the orientational order parameter $Q_4$.

\subsection{Relation between $\gamma$ and $Q_4$}

\begin{figure}[t]
\centering
\includegraphics[width=0.7\columnwidth]
{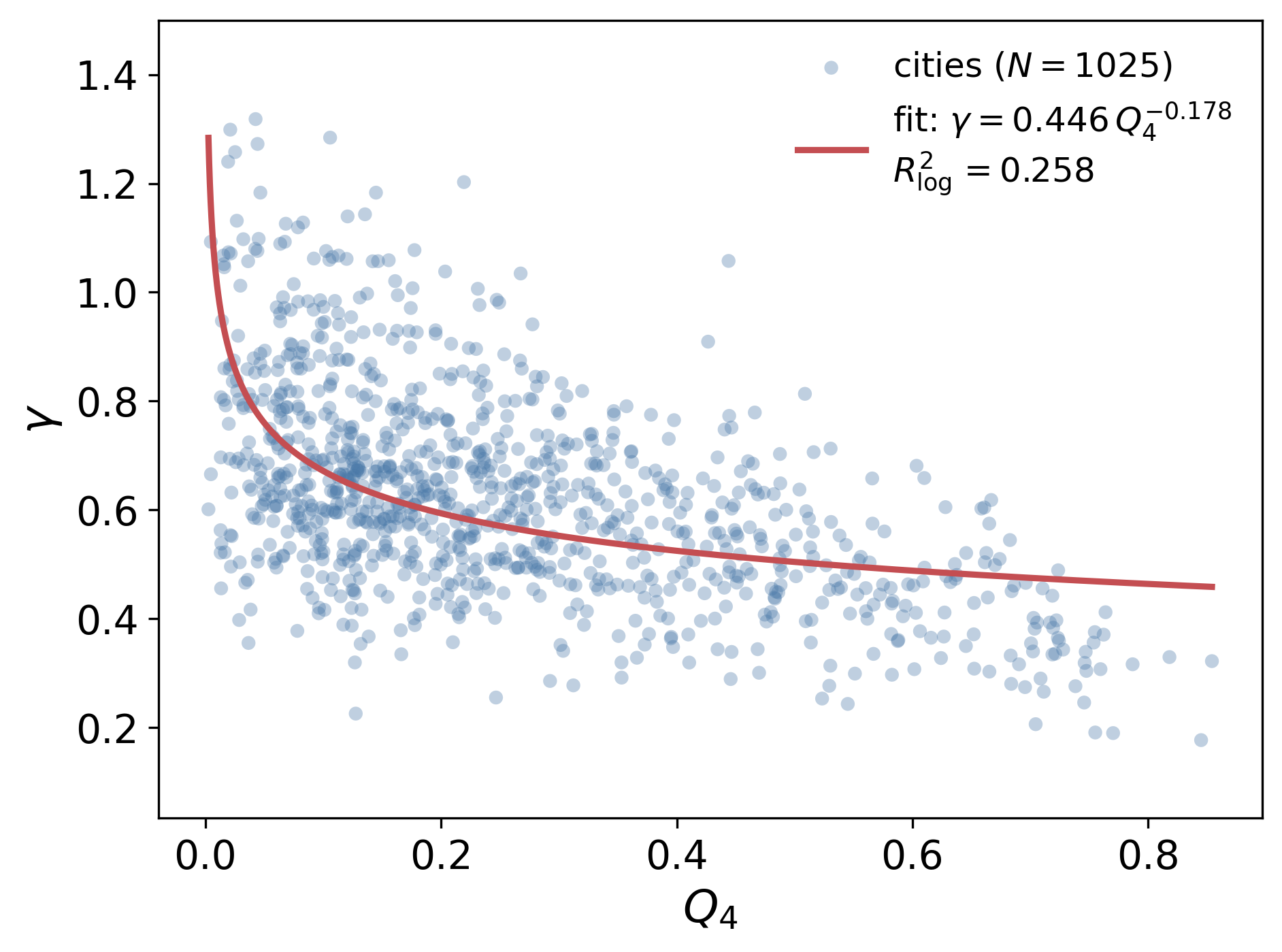}
\caption{Boundary irregularity $\gamma$ as a function of the
city-scale orientational order $Q_4$ for the $1025$ cities. The
solid line shows the power-law fit
$\gamma=AQ_4^\beta$, with $A=0.4458\pm0.0080$ and
$\beta=-0.1780\pm0.0094$.}
\label{fig:gamma_vs_Q4}
\end{figure}

We examine whether the geometrical irregularity of the bearings
communities, measured by $\gamma$, is related to the city-scale
orientational order $Q_4$. Across the $1025$ cities, $\gamma$ decreases
on average as $Q_4$ increases (Fig.~\ref{fig:gamma_vs_Q4}). A power-law
fit of the form
\begin{equation}
\gamma = A Q_4^{\beta}
\end{equation}
gives
\begin{equation}
A = 0.4458 \pm 0.0080,
\qquad
\beta = -0.1780 \pm 0.0094 .
\end{equation}
The negative exponent indicates that cities with stronger global street
orientation tend to have smoother community boundaries, whereas cities
with weak global order tend to display more irregular and interlocked
grains. The association is highly significant
($p=2.1\times10^{-68}$), but the dispersion remains large
($R^2=0.258$ in logarithmic space). Thus, global orientational order
accounts for only part of the variation in boundary geometry: cities
with comparable values of $Q_4$ can still have substantially different
values of $\gamma$.

\subsection{Relation between $\gamma$ and $N_{\mathrm{eff}}$}

\begin{figure}[t]
\centering
\includegraphics[width=0.7\columnwidth]
{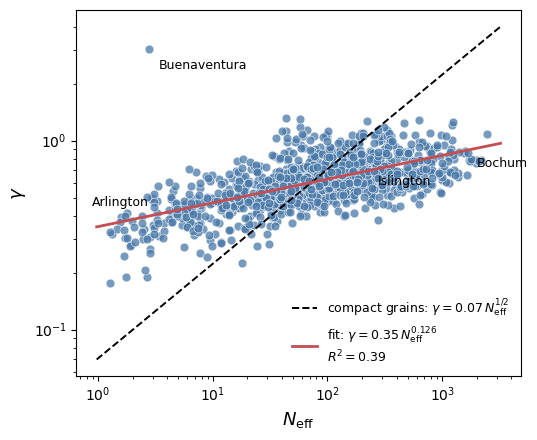}
\caption{Fragmentation measure $\gamma$ as a function of the
effective number of bearings communities $N_{\mathrm{eff}}$. The
solid line shows the power-law fit, with exponent
$\beta=0.126\pm0.010$ and $R^2=0.393$ in logarithmic space. The
line of slope $1/2$, with prefactor $c=0.071$, is shown only as a
visual reference for compact grains and is not a fit to the data.}
\label{fig:gamma_vs_Neff}
\end{figure}

We next examine how the geometrical fragmentation measured by $\gamma$
varies with the effective number of bearings communities
$N_{\mathrm{eff}}$. The two quantities are positively correlated
(Fig.~\ref{fig:gamma_vs_Neff}): cities containing a larger effective
number of grains tend, on average, to have more irregular and
interlocked community boundaries.

A power-law fit of the form
\begin{equation}
\gamma \sim N_{\mathrm{eff}}^{\beta}
\end{equation}
gives
\begin{equation}
\beta = 0.126 \pm 0.010 ,
\end{equation}
where the uncertainty denotes an approximate $95\%$ confidence
interval. The fit accounts for about $39\%$ of the variance in
logarithmic space ($R^2=0.393$). Thus, the number of grains is an
important determinant of fragmentation, although substantial
city-to-city variability remains.

For comparison, Fig.~\ref{fig:gamma_vs_Neff} also shows the dependence
\begin{equation}
\gamma_{\mathrm{comp}} =
c,N_{\mathrm{eff}}^{1/2},
\qquad c=0.071 ,
\end{equation}
expected for a collection of compact grains. The exponent $1/2$ is
shown only as a visual reference and is not fitted to the data. The much
smaller empirical exponent indicates that fragmentation increases more
slowly with $N_{\mathrm{eff}}$ than this compact-grain reference.

\section{Domain size, coherence, and the nematic order parameter}
\label{sm:nematic}

Throughout this section, ``domain'' means a bearings community retained
by the size filter used in the main text (at least five nodes), so that
the population described here is the same one entering
$\bar q_{\rm city}$, $N_{\rm eff}$ and $Q_4^{\mathrm{int}}$. We write
$\mathrm{int}(c)$ for the interior edges of domain $c$---those with both
endpoints in $c$---and
\begin{equation}
L_c \;=\; \sum_{e\,\in\,\mathrm{int}(c)} \ell_e
\label{eq:sm_Lc}
\end{equation}
for its internal street length. The domain-level order parameters
restated below are those defined in the main text; we repeat them here
so that this section can be read on its own.

\subsection{Domain sizes and the participation number}

The fourfold order of a domain is carried by the complex parameter
\begin{equation}
z_c \;=\; \frac{1}{L_c}\sum_{e\,\in\,\mathrm{int}(c)}
\ell_e\, e^{4i\theta_e} \;=\; q_c\, e^{4i\phi_c},
\label{eq:sm_zc}
\end{equation}
whose modulus $q_c\in[0,1]$ measures the internal coherence of the
domain and whose phase gives its dominant orientation $\phi_c$. Let
$N_c$ denote the number of edges in $\mathrm{int}(c)$---the edges
entering Eq.~(\ref{eq:sm_zc})---and $S_c$ the number of nodes of $c$.

Because $q_c$ is a \emph{length-weighted} resultant, the number of
independent contributions to it is not $N_c$ but the participation
number of the length weights,
\begin{equation}
N_{\rm eff}^{(c)} \;=\;
\frac{\left(\sum_{e\in\mathrm{int}(c)} \ell_e\right)^{2}}
     {\sum_{e\in\mathrm{int}(c)} \ell_e^{2}}
\;=\; \frac{L_c^{2}}{\sum_{e\in\mathrm{int}(c)} \ell_e^{2}} ,
\label{eq:sm_neff_c}
\end{equation}
the same inverse participation ratio used at the city level to define
$N_{\rm eff}$ in the main text: a domain built from a few long segments
averages less effectively than one built from many short ones. Across
the dataset the median ratio is $N_{\rm eff}^{(c)}/N_c=0.73$.

\begin{figure}
    \centering
    \includegraphics[width=0.49\textwidth]{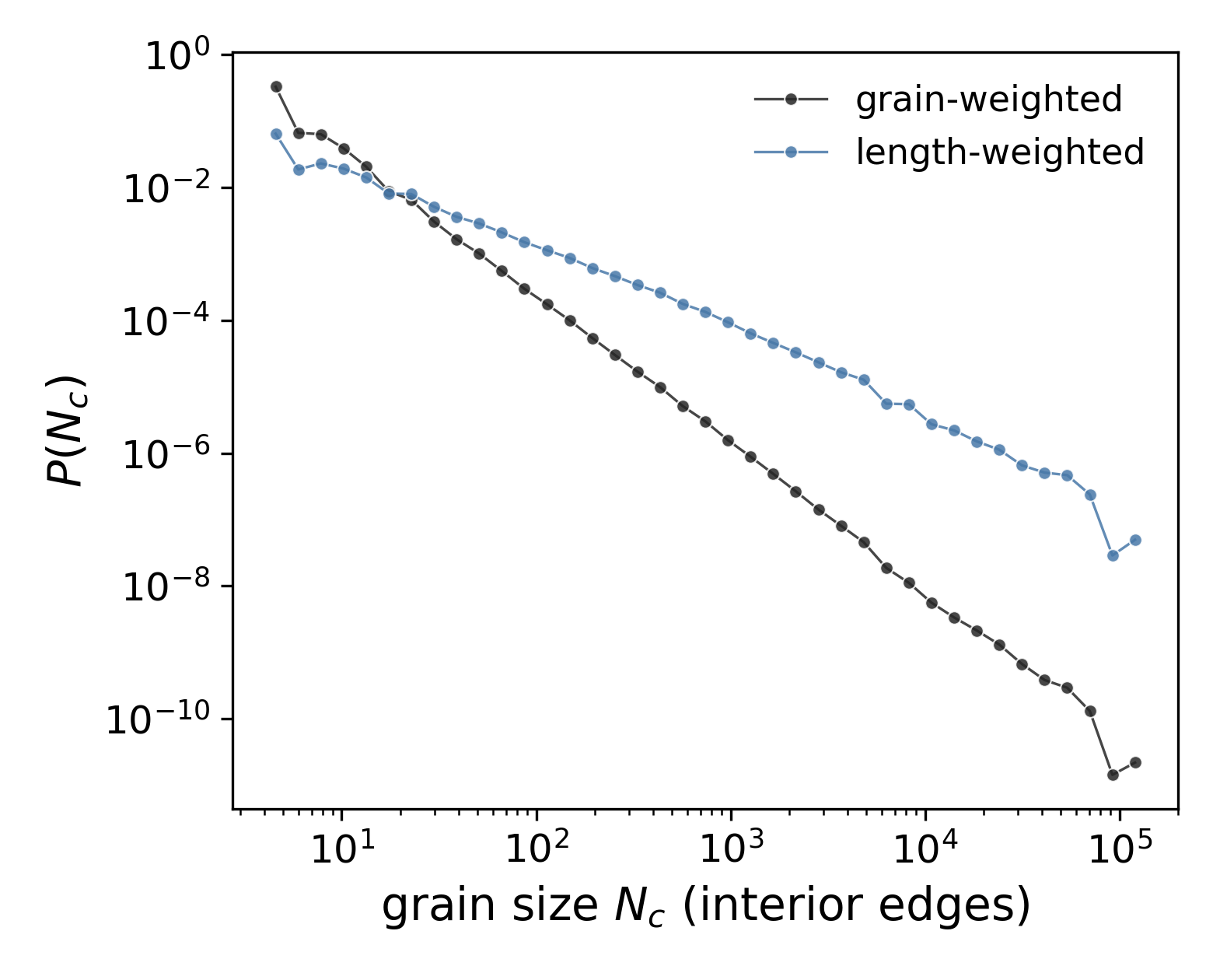}
\caption{Distribution of domain sizes $N_c$ (number of interior edges),
pooled over the $2.84\times10^{6}$ bearings communities of the $1025$
cities. Black: each domain counted once. Blue: each domain weighted by
its internal street length $L_c$. The two differ by more than an order
of magnitude in their median ($6$ against $74$), reflecting the very
large number of small tree-like domains that carry little street
length. The distribution extends over five decades with no
characteristic scale.}
    \label{smfig:Nc}
\end{figure}

The distribution of $N_c$ spans five decades (Fig.~\ref{smfig:Nc}). The median domain holds only $6$ interior edges on $7$ nodes and is therefore a tree. Since a connected subgraph
satisfies $N_c\ge S_c-1$, with equality only for a tree, the median
ratio $N_c/S_c=0.86$ implies that more than half of all domains are
trees: cul-de-sacs, branches and connector stubs, numerous but
individually tiny ($89\%$ but carrying $37\%$ of street length). These form the interstitial material between the
true crystallites, the analogue of the intergranular phase of a
nanocrystalline solid. Weighting by street length the median rises to
$N_c=74$ (interquartile range $[14,584]$), and street length is
distributed almost uniformly across half-decades of $N_c$ (third column
of Table~\ref{smtab:q2}), so that no characteristic domain size
emerges. At the opposite extreme, the largest single domain in the
dataset---in Chicago---contains $137\,610$ interior edges spanning
$4285$~km, that is $89\%$ of that city's street length, at $q_c=0.995$.

\subsection{The random-mosaic baseline for $q_c$}
\label{sec:sm_qrand}

To decide whether a measured coherence is large, it must be compared
with the value expected by chance. Consider a domain whose $N$ interior
edges carry independent bearings drawn uniformly on $[0,\pi)$. The
fourfold phases $4\theta_e$ are then uniform on $[0,2\pi)$ and the sum
in Eq.~(\ref{eq:sm_zc}) is a two-dimensional random walk of $N$ unit
steps. Its resultant length has mean $\tfrac{\sqrt\pi}{2}\sqrt{N}$, so
after normalisation by the number of steps the expected coherence is
\begin{equation}
\left\langle q \right\rangle_{\rm rand} \;=\;
\frac{\sqrt{\pi}}{2\sqrt{N}} .
\label{eq:sm_qrand_unit}
\end{equation}
When the edges are weighted by their lengths, the effective number of
independent steps is the participation
number~(\ref{eq:sm_neff_c}) rather than the raw count, and the baseline
becomes
\begin{equation}
\left\langle q \right\rangle_{\rm rand} \;=\;
\frac{\sqrt{\pi}}{2\,\sqrt{N_{\rm eff}^{(c)}}} .
\label{eq:sm_qrand}
\end{equation}
This is the domain-level analogue of the random-mosaic null used at the
city level in the main text: there the randomised objects are the
domain orientations $\phi_c$, here they are the individual edge
bearings within one domain. Equation~(\ref{eq:sm_qrand}) is evaluated
separately for every domain, using its own $N_{\rm eff}^{(c)}$.

\subsection{Two-axis fabric versus uniaxial alignment}

Because $e^{4i\theta}$ identifies $\theta$ with $\theta+\pi/2$, the
fourfold parameter~(\ref{eq:sm_zc}) cannot distinguish a two-axis grid
from a bundle of mutually parallel streets. The distinction is made by
the domain-level nematic parameter
\begin{equation}
q_c^{(2)} \;=\; \left|\frac{1}{L_c}
\sum_{e\,\in\,\mathrm{int}(c)} \ell_e\, e^{2i\theta_e}\right| ,
\label{eq:sm_q2c}
\end{equation}
which, for a domain whose dominant axis carries a length fraction $f$,
reduces to
\begin{equation}
q_c^{(2)} \;=\; \left|2f-1\right| ,
\label{eq:sm_q2f}
\end{equation}
and, for rectangular blocks of aspect ratio $a$, to
\begin{equation}
q_c^{(2)} \;=\; \frac{a-1}{a+1} .
\label{eq:sm_q2a}
\end{equation}

\begin{figure}
    \centering
    \includegraphics[width=0.98\textwidth]{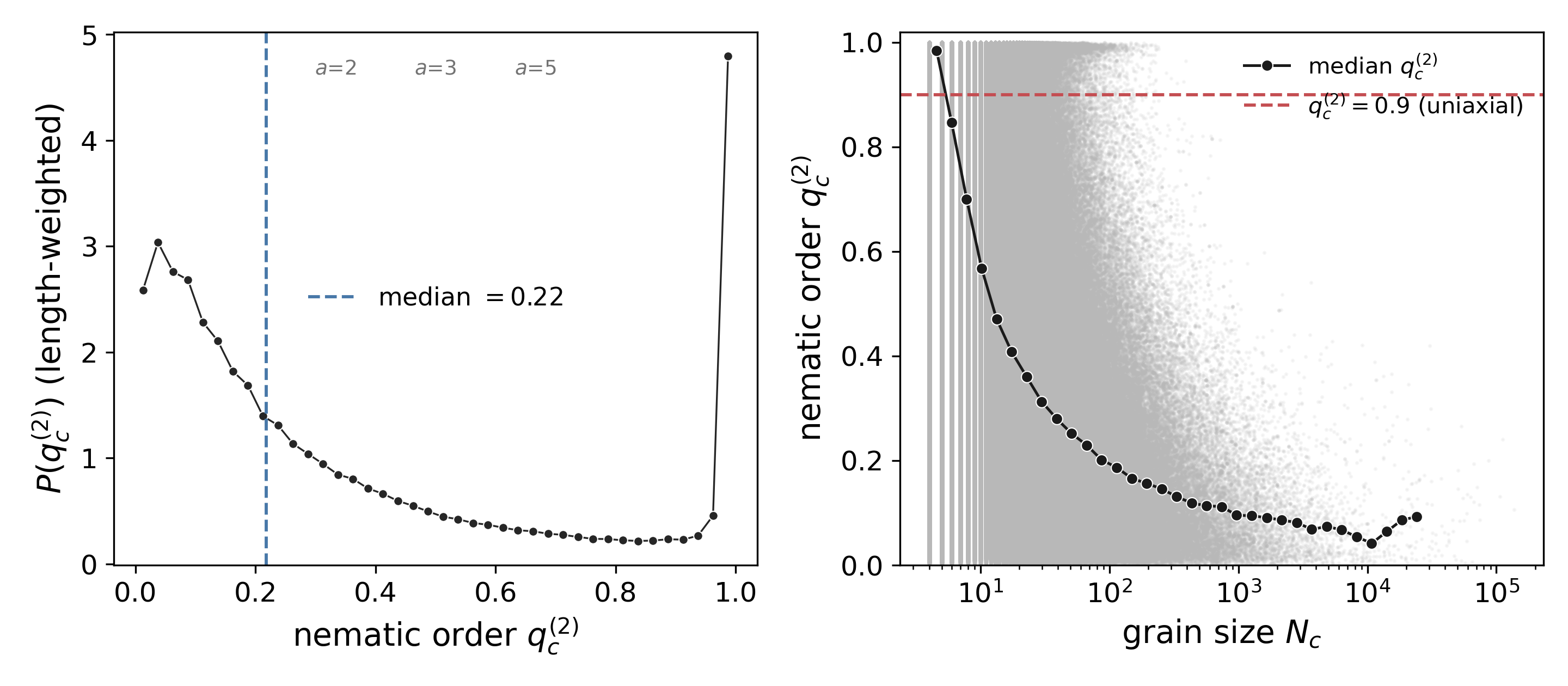}
\caption{Nematic order at the domain level. \textit{Left:}
length-weighted distribution of $q_c^{(2)}$ [Eq.~(\ref{eq:sm_q2c})]
pooled over all domains, with the median marked. Tick labels indicate
the block aspect ratios $a$ corresponding to
Eq.~(\ref{eq:sm_q2a}). The distribution is bimodal, the peak near unity
consisting of small tree-like domains that contain no blocks.
\textit{Right:} $q_c^{(2)}$ against domain size. Grey points are
individual domains, black symbols the median in logarithmic bins of
$N_c$. The dashed line at $q_c^{(2)}=0.9$ marks the uniaxial threshold.
Nematic order decreases monotonically with size: domains large enough
to contain blocks are two-axis fabrics with mild anisotropy.}
    \label{smfig:q2}
\end{figure}

Table~\ref{smtab:q2} resolves $q_c^{(2)}$ by domain size, and
Fig.~\ref{smfig:q2} shows the corresponding distributions. The uniaxial
population is confined to the smallest domains: those with
$q_c^{(2)}>0.9$ number $1.17\times10^{6}$ and carry $14.4\%$ of the
street length, but their median size is $N_c=5$ (ninetieth percentile
$10$), $99.5\%$ of them are trees, and $84\%$ of their length lies in
domains with $N_c<20$. A tree of five edges has no cycle and therefore
no block, so a second axis cannot exist within it; $q_c^{(2)}\simeq1$
there records the absence of a block structure rather than the presence
of a lamellar one. As soon as domains are large enough to contain
blocks the nematic order collapses: for $N_c\ge30$---$63.6\%$ of all
street length---the length-weighted median is $q_c^{(2)}=0.134$, an
aspect ratio $a\simeq1.31$ by Eq.~(\ref{eq:sm_q2a}), with only $2.0\%$
of that length uniaxial, and by $N_c\ge300$ the uniaxial fraction is
indistinguishable from zero.

The last two columns of Table~\ref{smtab:q2} record a separate and
stronger point. The median coherence $q_c$ is essentially independent
of domain size, varying only between $0.969$ and $0.977$ across five
decades, whereas the baseline~(\ref{eq:sm_qrand}) decays as
$(N_{\rm eff}^{(c)})^{-1/2}$. The ratio of the two therefore grows with
size, from $\simeq2$ in the smallest domains to $67$ in those with
$N_c>10^{3}$. A single pooled ratio ($5.2$, length-weighted over all
domains) understates the effect substantially, being dominated by the
very numerous small domains for which the baseline is weak.

The growth of the ratio is quantitatively that expected from a
size-independent coherence. Combining
Eqs.~(\ref{eq:sm_qrand}) and~(\ref{eq:sm_neff_c}) with $q_c\simeq0.97$
and $N_{\rm eff}^{(c)}/N_c\simeq0.73$ gives
\begin{equation}
\frac{q_c}{\langle q\rangle_{\rm rand}}
\;=\; \frac{2\,q_c}{\sqrt{\pi}}\sqrt{N_{\rm eff}^{(c)}}
\;\simeq\; 0.94\,\sqrt{N_c},
\label{eq:sm_ratio_pred}
\end{equation}
which reproduces the last column of Table~\ref{smtab:q2} to within
about $10\%$ over the whole range.

\begin{table}
\centering
\caption{Domain statistics resolved by size. $N_c$ is the number of
interior edges; the length share is the fraction of total street length
in each size class; $q_c^{(2)}$ and $q_c$ are length-weighted medians
within the class; the last column is the length-weighted mean of
$q_c/\langle q\rangle_{\rm rand}$, with $\langle q\rangle_{\rm rand}$
evaluated per domain from Eq.~(\ref{eq:sm_qrand}). Street length is
spread almost uniformly over size classes; nematic order falls
monotonically with size while coherence does not, so that the excess
over the random baseline grows as $\sqrt{N_c}$, in agreement with
Eq.~(\ref{eq:sm_ratio_pred}).}
\label{smtab:q2}
\begin{tabular}{lrrrrrr}
\toprule
$N_c$ & domains & length & median & frac.\ & median & ratio \\
      &         & share  & $q_c^{(2)}$ & $q_c^{(2)}>0.9$ & $q_c$ & to null \\
\midrule
$<10$          & $1\,958\,699$ & $0.177$ & $0.919$ & $0.510$ & $0.977$ & $2.2$  \\
$10$--$30$     & $651\,085$    & $0.188$ & $0.453$ & $0.217$ & $0.971$ & $3.5$  \\
$30$--$100$    & $175\,540$    & $0.174$ & $0.263$ & $0.065$ & $0.969$ & $6.1$  \\
$100$--$300$   & $40\,856$     & $0.137$ & $0.170$ & $0.009$ & $0.969$ & $10.8$ \\
$300$--$1000$  & $11\,652$     & $0.126$ & $0.120$ & $0.000$ & $0.969$ & $18.9$ \\
$\ge 1000$     & $3\,328$      & $0.199$ & $0.075$ & $0.000$ & $0.975$ & $67.1$ \\
\bottomrule
\end{tabular}
\end{table}

\bibliography{bibfile_street_network}